\documentclass[acmsmall,nonacm]{acmart}

\AtBeginDocument{%
  \providecommand\BibTeX{{%
    \normalfont B\kern-0.5em{\scshape i\kern-0.25em b}\kern-0.8em\TeX}}}

\setcopyright{acmcopyright}
\copyrightyear{2024}
\acmYear{2024}
\acmISBN{}
\acmDOI{}

\acmConference[arXiv '24]{arXiv}{Janurary, 2024}{Virtual}
\usepackage{comment}
\usepackage{algorithm}
\usepackage{algpseudocode}
\usepackage{graphicx}
\usepackage{textcomp}
\usepackage{xcolor}
\usepackage[switch]{lineno}
\usepackage{listings}
\usepackage{enumitem} 
\usepackage{tikz}
\usepackage{multirow}
\usepackage{placeins}

\definecolor{maroon}{rgb}{0.58,0,0.82}

\newcommand*\circled[1]{\tikz[baseline=(char.base)]{
            \node[shape=circle,draw,inner sep=1pt,] (char) {#1};}}

\newcommand{\code}[1]{{\small{\texttt{#1}}}}

\begin{document}

\title{The Parallel Semantics Program Dependence Graph}


\author{Brian Homerding}
\affiliation{%
  \institution{Northwestern University}
  \streetaddress{633 Clark St}
  \city{Evanston}
  \state{Illinois}
  \country{USA}
  \postcode{60208}
}
\email{brianhomerding2026@u.northwestern.edu}

\author{Atmn Patel}
\affiliation{%
  \institution{Northwestern University}
  \streetaddress{633 Clark St}
  \city{Evanston}
  \state{Illinois}
  \country{USA}
  \postcode{60208}
}
\email{atmn@u.northwestern.edu}

\author{Enrico Armenio Deiana}
\affiliation{%
  \institution{Northwestern University}
  \streetaddress{633 Clark St}
  \city{Evanston}
  \state{Illinois}
  \country{USA}
  \postcode{60208}
}
\email{enricodeiana2020@u.northwestern.edu}

\author{Yian Su}
\affiliation{%
  \institution{Northwestern University}
  \streetaddress{633 Clark St}
  \city{Evanston}
  \state{Illinois}
  \country{USA}
  \postcode{60208}
}
\email{yiansu2018@u.northwestern.edu}

\author{Zujun Tan}
\affiliation{%
  \institution{Princeton University}
  \streetaddress{330 Alexander St}
  \city{Princeton}
  \state{New Jersey}
  \country{USA}
  \postcode{08540}
}
\email{zujunt@cs.princeton.edu}

\author{Ziyang Xu}
\affiliation{%
  \institution{Princeton University}
  \streetaddress{330 Alexander St}
  \city{Princeton}
  \state{New Jersey}
  \country{USA}
  \postcode{08540}
}
\email{ziyangx@princeton.edu}

\author{Bhargav Reddy Godala}
\affiliation{%
  \institution{Princeton University}
  \streetaddress{330 Alexander St}
  \city{Princeton}
  \state{New Jersey}
  \country{USA}
  \postcode{08540}
}
\email{bgodala@cs.princeton.edu}

\author{David I. August}
\affiliation{%
  \institution{Princeton University}
  \streetaddress{330 Alexander St}
  \city{Princeton}
  \state{New Jersey}
  \country{USA}
  \postcode{08540}
}
\email{august@princeton.edu}

\author{Simone Campanoni}
\affiliation{%
  \institution{Northwestern University}
  \streetaddress{633 Clark St}
  \city{Evanston}
  \state{Illinois}
  \country{USA}
  \postcode{60208}
}
\email{simone.campanoni@northwestern.edu}

\renewcommand{\shortauthors}{Homerding, et al.}

\begin{abstract}
  A compiler's {\em intermediate representation (IR)} defines a program's execution plan by encoding its instructions and their relative order. 
Compiler optimizations aim to replace a given execution plan (which instructions to execute and when) with a semantically-equivalent one that increases the program's performance for the target architecture.
Alternative representations of an IR, like the Program Dependence Graph (PDG), aid this process by capturing the minimum set of constraints that semantically-equivalent execution plans must satisfy.
Parallel programming like OpenMP extends a sequential execution plan by adding the possibility of running instructions in parallel, creating a parallel execution plan.
Recently introduced parallel IRs, like TAPIR, explicitly encode a parallel execution plan.
These new IRs finally make it possible for compilers to change the parallel execution plan expressed by programmers to better fit the target parallel architecture.
Unfortunately, parallel IRs do not help compilers in identifying the set of parallel execution plans that preserve the original semantics. 
In other words, we are still lacking an alternative representation of parallel IRs to capture the minimum set of constraints that parallel execution plans must satisfy to be semantically-equivalent. 
Unfortunately, the PDG is not an ideal candidate for this task as it was designed for sequential code.
In more detail, this paper shows that the PDG over-constrains the optimization space when used for parallel code.
We propose the Parallel Semantics Program Dependence Graph (PS-PDG) to precisely capture the salient program constraints that all semantically-equivalent parallel execution plans (and therefore parallel IRs) must satisfy.
This paper defines the PS-PDG, justifies the necessity of each extension to the PDG, and demonstrates the increased optimization power of the PS-PDG over an existing PDG-based automatic-parallelizing compiler.
Compilers can now rely on the PS-PDG to select different parallel execution plans while maintaining the same original semantics.

\end{abstract}

\maketitle

\section{Introduction}
A compiler's {\em intermediate representation (IR)} defines a program's execution plan by encoding its instructions and their relative order.
Most compiler optimizations are performed by changing the IR of a program. 
Many of these optimizations (e.g., code scheduling) need to change the relative order of IR instructions (i.e., the execution plan) to reach their optimization goal and/or to better target the underlying architecture.
This is possible because not all orders specified by a given IR instance are necessary (thanks to having independent instructions).
It is therefore important to understand what is the minimum \emph{subset} of instruction order constraints that must be enforced to preserve the original semantics.
Unfortunately, IRs (e.g., LLVM IR) do not highlight such subset; instead, a specific instance of an IR specifies the \emph{total} order of its instructions, but some of these orders are the result of a choice (e.g., execution order of independent instructions within a basic block) rather than a constraint that must be satisfied.
To overcome this limitation, code optimizations rely on a different representation of the IR code called the \emph{Program Dependence Graph} (PDG)~\cite{pdg}, which encodes the (ideally small) set of order constraints that all possible total order of instructions (hence, IR instances) must satisfy to preserve the original semantics of the input code.
The PDG can alternatively be seen as the (ideally large) set of degrees of freedom that a code optimizer can leverage to generate an IR instance that better fits the target architecture.

\begin{figure}[h]
\centering
\includegraphics[keepaspectratio, width=.90\textwidth]{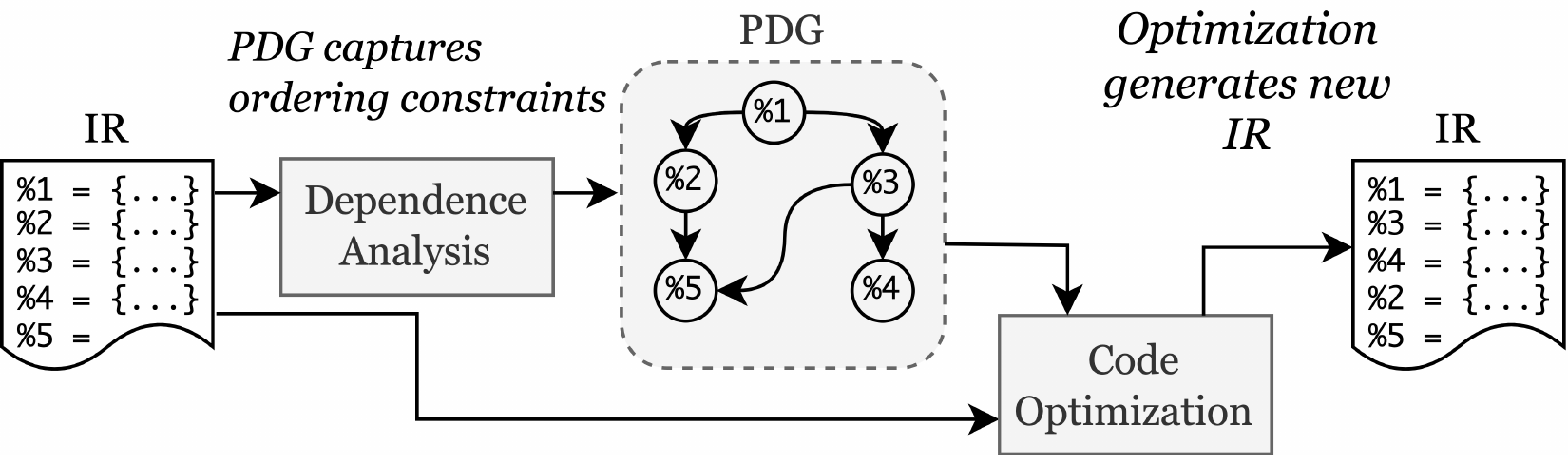}
\caption{Compiler pipeline utilizing IR and PDG for code optimization. }
\label{fig:intro_seq}
\end{figure}

The typical compilation pipeline followed by a code optimization is shown in Figure~\ref{fig:intro_seq}.
A given IR instance is analyzed by a dependence analysis to generate its PDG~\footnote{Some code optimizations target small code regions like loops. Some compilers therefore generate only the subset of the PDG that is needed by the selected code optimizations.}.
Then, a code optimization decides the order of instructions that are left unspecified by the PDG (those that can execute in multiple orders).
These decisions are then enforced by generating a new IR instance with (potentially) a different total order compared to the one given as input.
In other words, a code optimization changes the execution plan of an IR by relying on the degrees of freedom highlighted by its PDG representation.

For decades, compilers have used sequential IRs (IRs that assume sequential execution; e.g., LLVM IR, GCC RTL) and PDG to compile and optimize both sequential and parallel programs.
Parallel code extends the execution plan of sequential code by adding the choice of which instructions to run in parallel with respect to what.
Parallel language compilers (e.g., OpenMP compilers) generally represent the parallel aspects of the execution plan (e.g., run iterations of a given loop in parallel) with annotations, built-ins, or calls added to their IRs.
These constructs enable sequential optimizations to proceed correctly without changing the parallel aspects of the execution plan.
This is obtained by outlining parallel regions into functions so that the sequential interpretation of the resulting code is valid (even if overly constrained).
This design decision handcuffed compilers to maintain the parallel aspects of the execution plan that were encoded by programmers.
This is the reason why OpenMP and Cilk compilers (for example) do not alter the decision of what to run in parallel that was encoded by programmers within their source code.

While sequential IRs and their PDG made sense when compilers were not expected to include code optimizations specifically designed for parallel code, they are insufficient for the next generation of compilers that include parallel optimizations.
This is because parallel optimizations require modifying parallel aspects of the input code.
In more detail, as parallel machines proliferate, the community frequently observed that the parallel execution plan expressed by programmers is sub-optimal for the target architecture~\cite{sarkar_thesis,pp_diverse_arch,pp_herterogeneous_arch,raja,pp_data_par_lang,kremlin,pp_kernel_synth}.
This is because different architectures require different parallel execution plans to reach high performance.
In other words, just as the sequential execution plans specified by programmers are sub-optimal~\cite{lam1988software,DBLP:conf/isca/FanCJ19,tran2017clairvoyance,Reddi:2010:EVE:1839667.1839674,lee2000compiler,10.1007/978-3-642-02737-6_12,Fan:2018:CIC:3195970.3196013}, so too are the parallel execution plans found in manually-written parallel programs.
To address this, the community has introduced parallel IRs like TAPIR~\cite{tapir,ppg,sarkar_opt_ppg,wolfe_pcfg,llvm-omp,inspire,psg} to represent parallel execution plans and changes thereto.
Explicit encoding of a specific parallel execution plan into a parallel IR made changes to it finally possible.
Unfortunately, changing a given parallel execution plan is still not practical because a given instance of a parallel IR does not highlight the minimum set of constraints that all parallel execution plans must satisfy to preserve the original semantics.
Similarly to the need to have a separate representation for sequential IRs to enable code optimizers to change the sequential execution plan (i.e., the total order of instructions), we need a separate representation of parallel IRs to enable parallel code optimizers \emph{to change the parallel execution plan} (i.e., what to run in parallel and how).
Unfortunately, the PDG is not a good representation for this goal because it was designed to target sequential IRs, which blocks it from encoding all degrees of freedom that parallel code optimizers could take advantage of.
In other words, the PDG is a sub-optimal representation for parallel IRs and our empirical results (\S\ref{sec:system_methodology}) clearly show that its limitations for parallel IRs are too significant to be ignored.

To overcome this limitation, this paper proposes the Parallel Semantics Program Dependence Graph (PS-PDG) representation to capture the salient program constraints of parallel programs.  
The PS-PDG generalizes the PDG to capture the constraints that define the set of semantically equivalent execution plans for modern parallel programming models such as OpenMP and Cilk.
After defining the PS-PDG, this paper shows how each of its extensions to the PDG is necessary.

\begin{figure}[h]
\centering
\includegraphics[keepaspectratio, width=.90\textwidth]{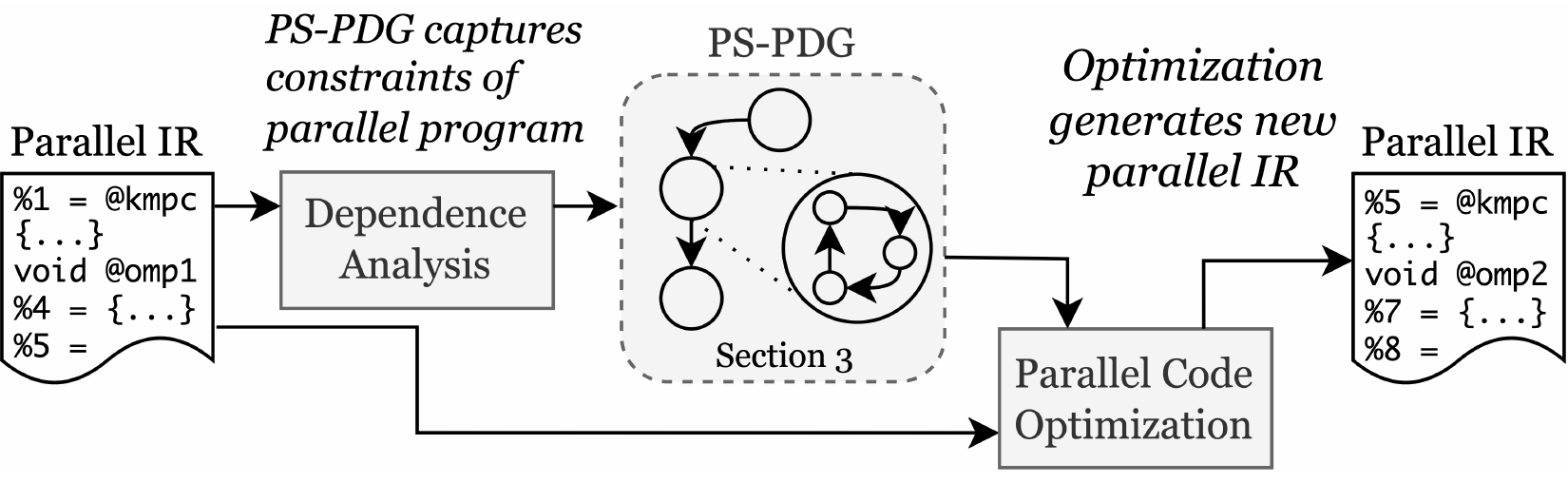}
\caption{Compiler pipeline utilizing parallel IR and PS-PDG for parallel code optimization. }
\label{fig:intro_par}
\end{figure}

To reduce disruptions to existing compilers, we propose a PS-PDG-based compilation pipeline (shown in Figure~\ref{fig:intro_par} and described next) that is similar to what has been in use for decades.
Our pipeline substitutes the PDG representation with the PS-PDG one.
We believe this is the missing piece to enable parallel compilers to reach their full potential.
In more detail, in a compiler developed with the PS-PDG, Cilk\cite{cilk} and OpenMP\cite{openmp50} source code is first translated into their parallel IR while preserving the execution plan expressed by the programmers.
Then, the IR is analyzed to generate a PS-PDG, which captures the minimum constraints necessary to preserve the original semantics.
With the PS-PDG, the compiler is finally capable to identify all possible parallel execution plans that are guaranteed to preserve the original semantics of the input code.
Hence, parallel compilers can now find the most appropriate parallel execution plan for the target architecture.
The parallel execution plan chosen is then realized into the generated parallel IR.
The parallel IR is then translated into the target assembly code.

The main contributions of this work are:
\begin{itemize}
  \item A definition of the PS-PDG: a representation that captures the salient program constraints of parallel programs (\S\ref{sec:pspdg}).
  \item A detailed analysis of the necessity of each element of the PS-PDG representation (\S\ref{sec:theoretical_evaluation}).
  \item A demonstration that the PS-PDG is sufficient to fully capture the parallel semantics of OpenMP programs (\S\ref{sec:sufficiency}).
  \item The first compiler to generate the PS-PDG for existing OpenMP benchmarks (\S\ref{ssec:impl}).
  \item A demonstration of PS-PDG's ability to expand the options available to an automatic-parallelizing compiler when selecting a parallelization plan. We compare this to the PDG and source code parallelization plan (\S\ref{ssec:space}).
  \item A measurement of the reduction in the critical path on an ideal machine using existing automatic-parallelizing compiler techniques with the PS-PDG. We compare this to the PDG and source code parallelization plan (\S\ref{ssec:speedup}).   
\end{itemize}


\section{Background \& Motivation}
Programmers need to express well-tuned parallelism in their applications to achieve high performance and energy efficiency.
This requires rich parallel programming models (PPMs) so that programmers can express complex parallel execution plans for their applications.
This has led to an ever-increasing number of PPMs and features therein.
Two examples of widely-adopted PPMs are OpenMP and Cilk.
This section uses an OpenMP code example to demonstrate how a programmer can explicitly encode a parallel execution plan.
Then, the same example serves as motivation for using the PS-PDG to represent the precise parallel constraints of a parallel program for use in parallel execution plan transformations.

\subsection{Programmers Explicitly Encode Parallelism}
The OpenMP PPM allows programmers to parallelize their code with pragmas.
An example is \code{\#pragma omp parallel for} which specifies that the iterations of the annotated loop can execute in parallel.
In this pragma, the worksharing pragma \code{omp for} specifies that the iterations of the loop should be distributed across multiple threads.
Other pragmas offer greater control over the parallel execution, such as \code{critical} which specifies that the given code region should only be executed by a single thread at any given time.

\begin{figure*}[t]
\centering
\includegraphics[keepaspectratio, width=.98\textwidth]{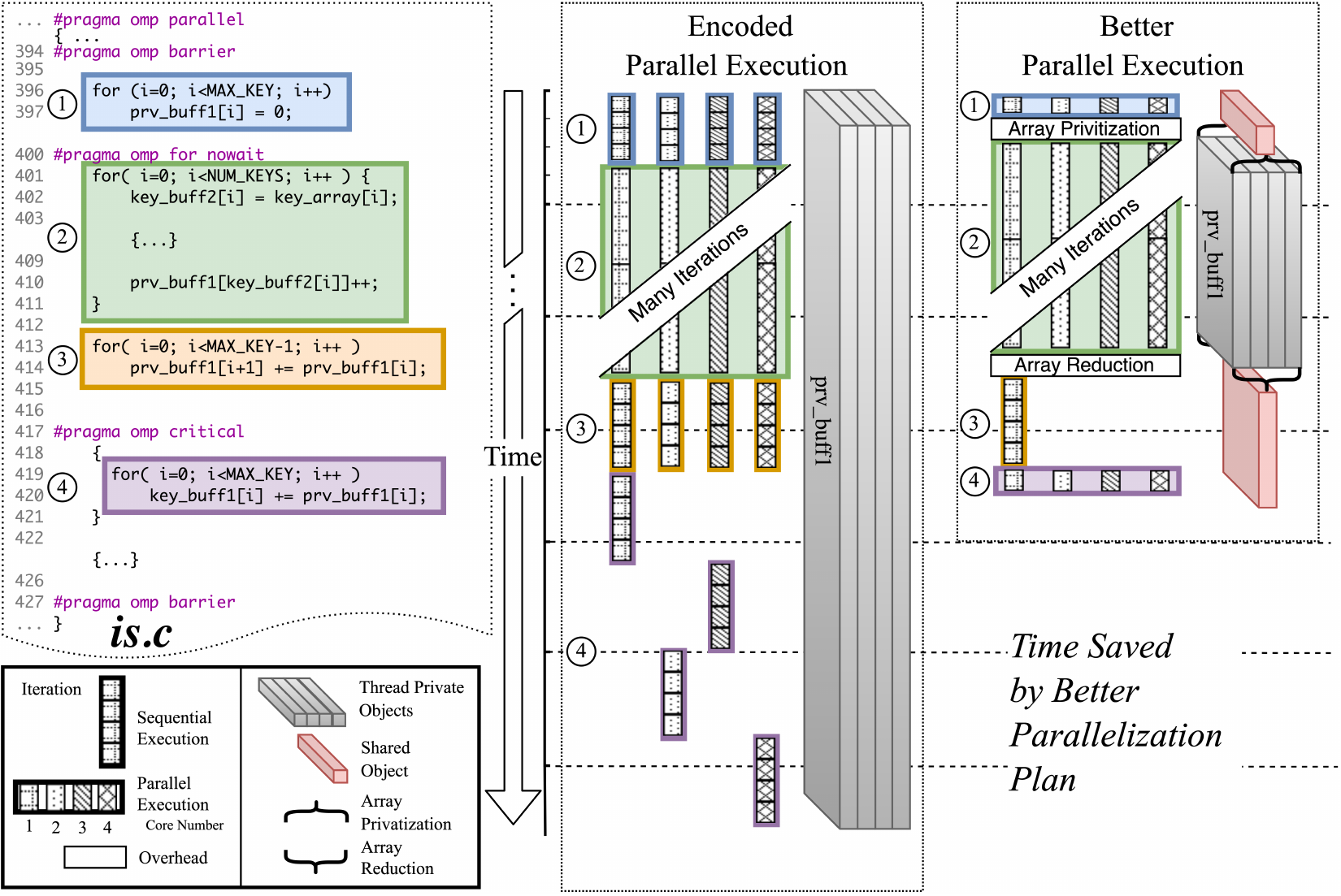}
\caption{The key computational kernel from the IS benchmark with the original and a more performant compiler-selected parallel execution.}
\label{fig:motivation_original}
\end{figure*}

Fig.~\ref{fig:motivation_original} shows the OpenMP source code from the hottest computation kernel of the \code{IS} benchmark from the \code{NAS} benchmark suite~\cite{nas}, along with the execution of its encoded parallelization plan.
The entire kernel is within a \code{\#pragma omp parallel} which spawns many threads.  
Since the loops \circled{1} (blue) and \circled{3} (orange), do not have a worksharing pragma, each thread executes the entire loop on its private copy of \code{prv\_buff1}.
Loop \circled{2} (green) instead has its iterations running in parallel between threads.
Loop \circled{4} (purple) is wrapped in a \code{critical} section to avoid a data race while updating \code{key\_buff1} concurrently. 

The code of Fig.~\ref{fig:motivation_original} is an example of the OpenMP PPM where the programmer has encoded a specific parallelization plan into the application.
A {\em parallelization plan} is the selection of what to parallelize (e.g., which loops) along with enabling features (e.g., which variables are thread-private), combined with the chosen parallel execution model (e.g., tasks, threads).  
Beyond the explicit parallelization encoded, the parallelization plan implies properties of the original code.

\label{sec:motivation}

\subsection{Enabling Compilers to Optimize Parallel Code}
Let us re-consider the hot code of \code{IS} shown in Fig.~\ref{fig:motivation_original}, but with a different parallelization plan than the one encoded by its OpenMP implementation, shown on the right side of Fig.~\ref{fig:motivation_original}.
Now iterations of loop \circled{1} execute in parallel while accessing different slices on a shared copy of \code{prv\_buff1} (the original plan had loop \circled{1} accessing only the thread-private copies of this array).
We divide the iterations of loop \circled{1} across many threads that access different slices of the single shared array \code{prv\_buff1}.
Then, we perform an array privatization of \code{prv\_buff1} before executing loop \circled{2}.
Iterations of loop \circled{2} execute between threads as in the original plan.  We then reduce the private copies of \code{prv\_buff1} to their shared copy as soon as loop \circled{2} ends its parallel execution.
Therefore, loop \circled{3} will only need to execute on a single thread (avoiding its parallel overhead that the original plan had).
As there is only one shared copy of \code{prv\_buff1} at this point of execution, loop \circled{4} can now execute in parallel (the original plan executed the loop sequentially between threads) by dividing its iterations between threads (without any critical section).
This new parallelization plan's execution is shown on the right side of Fig.~\ref{fig:motivation_original}.

\begin{figure*}[!t]
\centering
\includegraphics[keepaspectratio, width=.98\textwidth]
{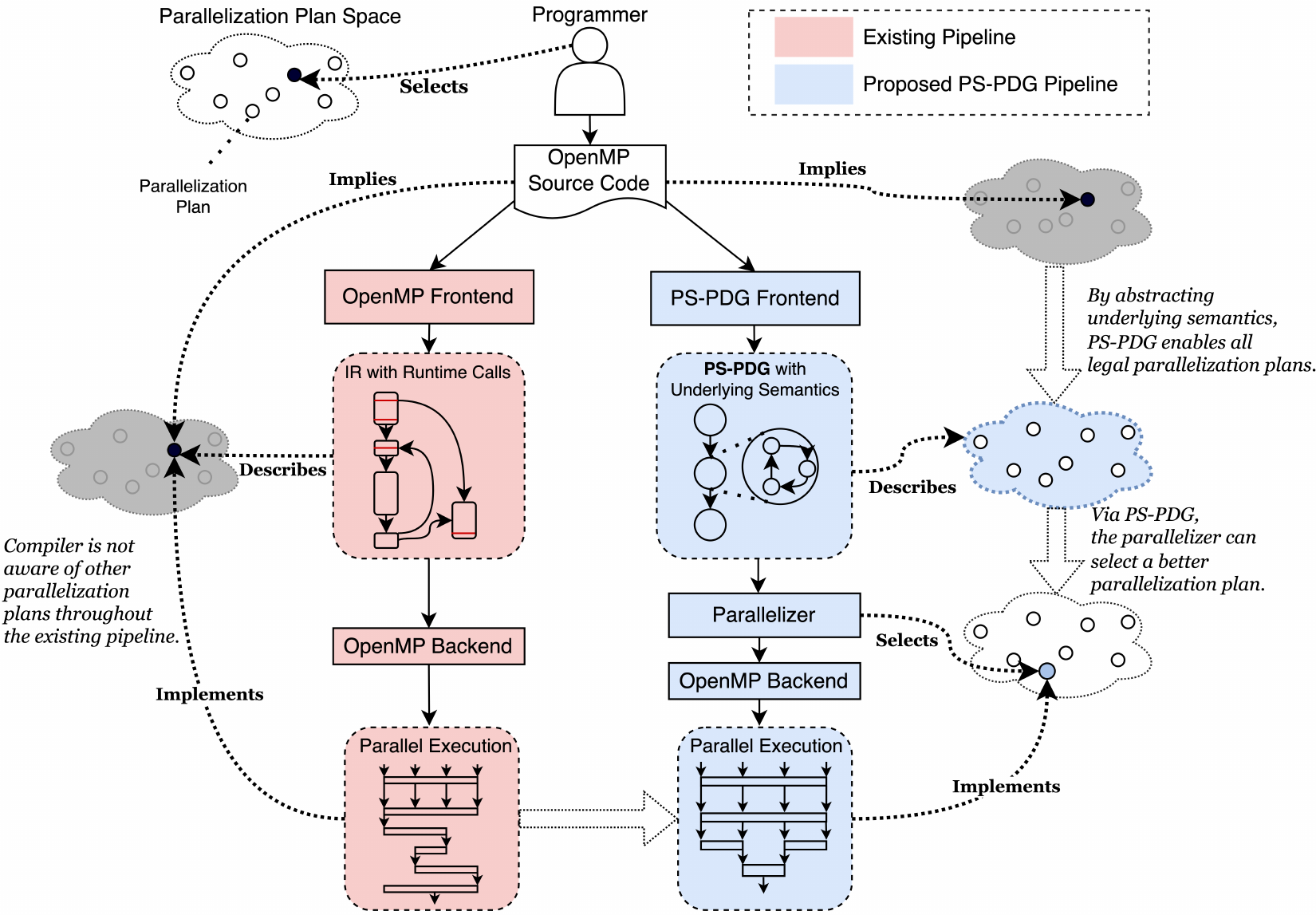}
\caption{Comparison of the existing pipeline with our proposed PS-PDG pipeline. The parallelization plan in the source code is abstracted away.}
\label{fig:motivation_ideal_pipe}
\end{figure*}

Consider the transformations required to change the parallelization plan on the left to the parallelization plan on the right of Fig.~\ref{fig:motivation_original}.
The private buffer \code{prv\_buff1} of loop \circled{2} must first be recognized as private from the IR explicit representation, then its uses beyond the loop must be considered.
The updates to \code{prv\_buff1} in both loops \circled{2} and \circled{3} must be recognized as reducible in both uses before either can be transformed.
Furthermore, the compiler must leverage the developer knowledge that the various arrays do not alias with one another and that the indirect index into \code{prv\_buff1} does not go beyond the use of the \code{prv\_buff1} in the other loops.

This example suggests that compilers need to become capable of modifying the parallelization plan expressed by programmers.
To do so, compilers must be equipped with the correct abstraction that captures the precise parallel constraints of a parallel program.
Today's compilers of PPMs like OpenMP and Cilk cannot do it.
The compilation pipeline of these compilers look like the one shown on the left of Fig.~\ref{fig:motivation_ideal_pipe}.
As the source code is processed by the compiler, the parallelization plan is simply lowered to the runtime calls that implement it; these compilers do not have an abstraction that captures the precise parallel constraints of the parallel program being compiled.
To empower compilers to transform the parallelization plan to better utilize the underlying hardware, we need to change the compiler's internal abstractions.

To perform transformations such as the one described above, the compiler needs to implement the transformed parallelization. 
Today's automatic parallelizing compilers~\cite{helix,dswp,matni2022noelle,perspective,scaf} successfully use the PDG abstraction, but they cannot rely on the parallel semantics expressed by programmers because the PDG does not capture it.
To overcome this limitation, this paper proposes the PS-PDG, an abstraction capable of capturing the precise parallel constraints of parallel programs while decoupling it from the encoded parallel execution plan.
With the PS-PDG, compilers can now explore the space of semantically equivalent parallelization plans, while preserving semantics, to fully leverage the precise parallel constraints of the parallel program.

The PS-PDG enables the pipeline shown in Fig.~\ref{fig:motivation_ideal_pipe}.
The proposed pipeline does not rigidly follow the encoded parallelization plan (like today's compilers do).
In this new pipeline, the precise parallel constraints of a parallel program is captured by the PS-PDG abstraction so that compilers can now see the space of semantically-equivalent parallelization plans.
The middle-end is now capable of selecting the parallelization plan that best fits the underlying architecture.

\section{PS-PDG Definition}
\label{sec:pspdg}
PPMs like OpenMP and Cilk enable programmers to make parallelization decisions explicit.
A programmer can decide where to spawn threads or tasks, to distribute the computation of a loop between parallel threads and/or tasks, and how to synchronize their execution.
These parallelization decisions define the \emph{parallel execution plan} of that program.

Beyond controlling what code can run in parallel and when it can do so, a parallel execution plan also \emph{implies properties of the code} of the original program.
For example, a parallel execution plan described using OpenMP can include the declaration that iterations of a loop will run in parallel during their executions.
This plan implies the property that the target loop has no loop-carried dependences between its iterations.
Another example is an OpenMP critical section in a loop, which implies both the need to enforce the atomic property of the target code segment and that any order of invocations of the target segment between loop iterations is valid.
We refer to this implied information as \textbf{the precise constraints of a parallel program}, which is captured by the PS-PDG.

\begin{table}[h]
    \caption{Complete PS-PDG Definition}
    \label{tab:definition}
    \centering
    \begin{tabular}{l l}

        \textbf{PS\text{-}PDG} & ::= (Node$^+$, Edge$^*$, Variable$^*$, VariableAccess$^*$) \\
        \textbf{Node} & ::= (Instruction, Trait$^*$) $|$ (HierarchicalNode, Trait$^*$) \\
        \textbf{HierarchicalNode} & ::= (Node$^+$, Context?) \\
        \textbf{Trait} & ::= (Singuler $|$ Unordered $|$ Atomic, Context) \\
        \textbf{Edge} & ::= DirectedEdge $|$ UndirectedEdge \\
        \textbf{DirectedEdge} & ::= (Node$_{\mathrm{producer}}$, Node$_{\mathrm{consumer}}$, Data\text{-}selector?) \\
        \textbf{UndirectedEdge} & ::= (Node, Node, Context) \\
        \textbf{Data\text{-}selector} & ::= (Any\text{-}Producer $|$ Last\text{-}Producer $|$ All\text{-}Consumers, Context) \\
        \textbf{Variable} & ::= (Privatizable $|$ Reducible, Context) \\
        \textbf{VariableAccess} & ::= (Variable, Node$_{\mathrm{use}}^*$, Node$_{\mathrm{def}}^*$) \\
        \textbf{Context} & ::= \textit{Unique Identifier}  \\ 
    \end{tabular}

\end{table}

The PS-PDG extends the PDG abstraction to capture the precise parallel constraints of a parallel OpenMP or Cilk program.
Like the PDG, the PS-PDG has nodes to represent computation and edges to represent dependences within the computation, but it also includes variables to represent data and use/def edges to represent the relation between data and its computation.  
As shown in Table~\ref{tab:definition}, a PS-PDG consists on one or more nodes with zero or more edges, variables and variable accesses.
The rest of this Section will describe each extension in detail.

\subsection{Hierarchical Nodes}
Explicit parallel programming enables programmers to specify properties of a code region.
Often such properties do not hold at finer granularities (e.g., single instruction).
For example, an OpenMP \emph{critical} section declares that the code region as a whole has the atomicity property.
This atomic property does not hold at a finer granularity like at the single instruction level that composes this critical code region.
For this reason, the PS-PDG both adds the ability to have a single node that represents an entire code region and the ability to express properties at their node granularity (\S\ref{sec:traits}).

\begin{figure}[h]
\centering
\includegraphics[keepaspectratio, width=.75\textwidth]{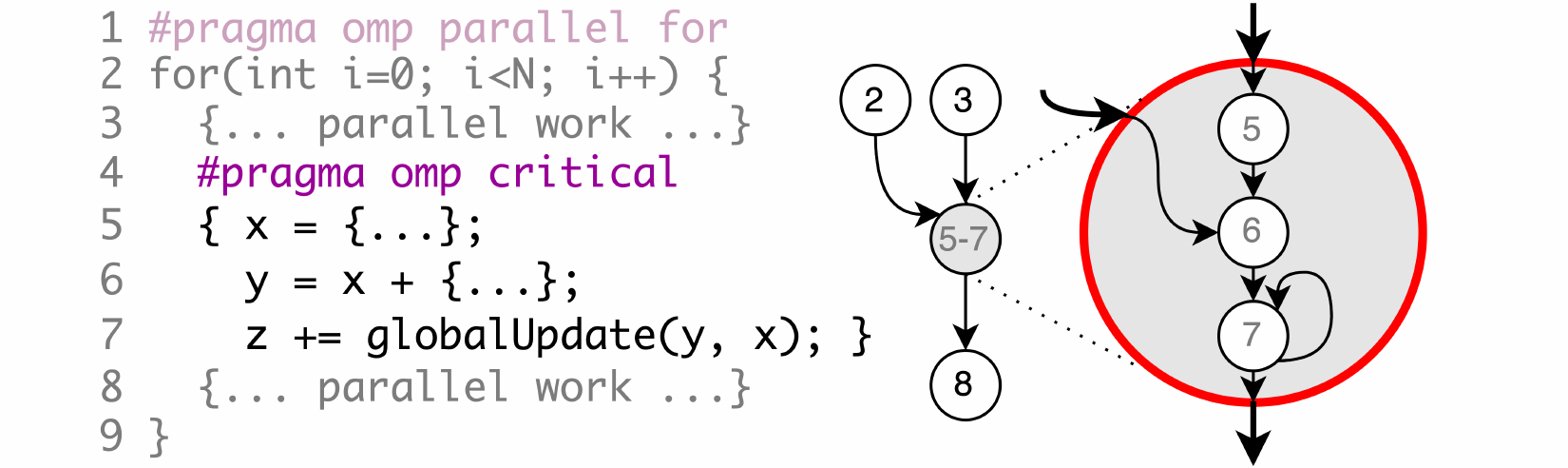}
\caption{Capturing properties of a region into hierarchical nodes with traits}
\label{fig:def_hierarchy}
\end{figure}

A node in the PS-PDG represents a non-empty set of instructions organizing the code hierarchically, shown in Fig.~\ref{fig:def_hierarchy}.
For example, all instructions of a \emph{critical} section in the PS-PDG is represented by a single node.
More generally, a node \emph{N} of the PS-PDG is a non-empty set of one or more instructions or other nodes such that both direct and indirect self-inclusions are not allowed.
Having a single node representing a set of instructions is needed to capture the parallel semantics of parallel constructs that target more than a single instruction.

\subsection{Node Traits}
\label{sec:traits}
Some properties expressed in a PPM are traits of a code region.
These traits can be important for the correctness and/or performance of a parallel application, for instance, atomicity.

\begin{figure}[h]
\centering
\includegraphics[keepaspectratio, width=.75\textwidth]{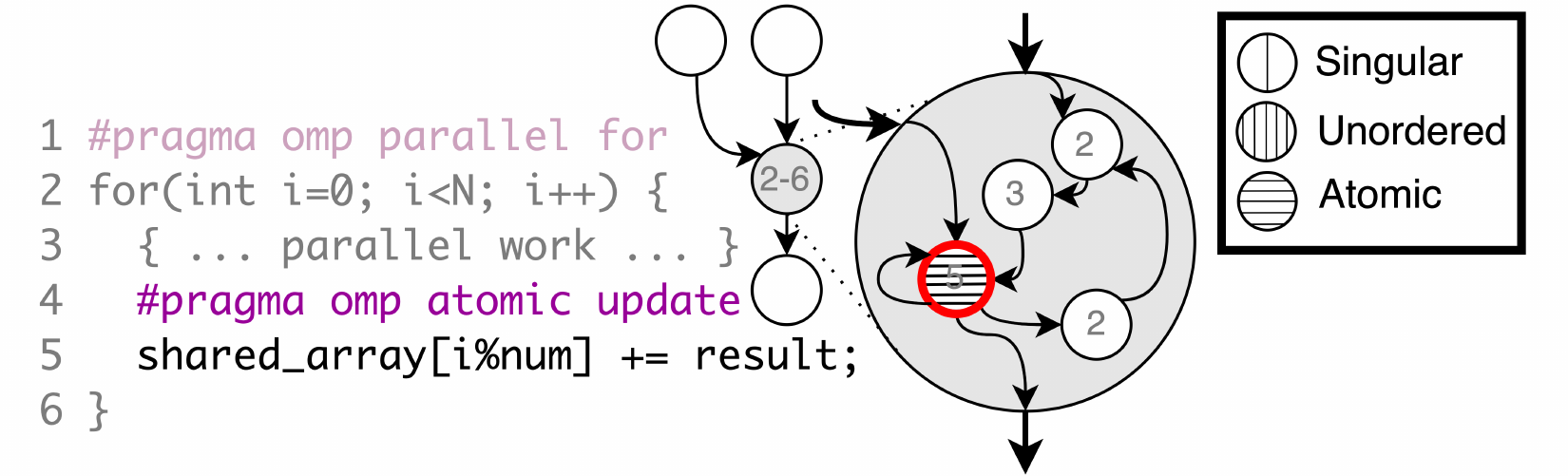}
\caption{How traits can be used to capture atomic updates.}
\label{fig:def_node_traits}
\end{figure}

A node in the PS-PDG can have various traits.
This paper implemented the three types of traits that are enough for the target languages OpenMP and Cilk: the atomic, orderless, and singular traits.
An atomic node represents a set of computations that must be executed atomically during its parallel execution.
An orderless node expresses that different instances of that node can be executed in any order for a given context.
A singular node represents a set of computations that must be executed by only a single instance for a given context.
An example of a node traits is shown in Fig.~\ref{fig:def_node_traits}.

\begin{figure}[b]
\centering
\includegraphics[keepaspectratio, width=.75\textwidth]{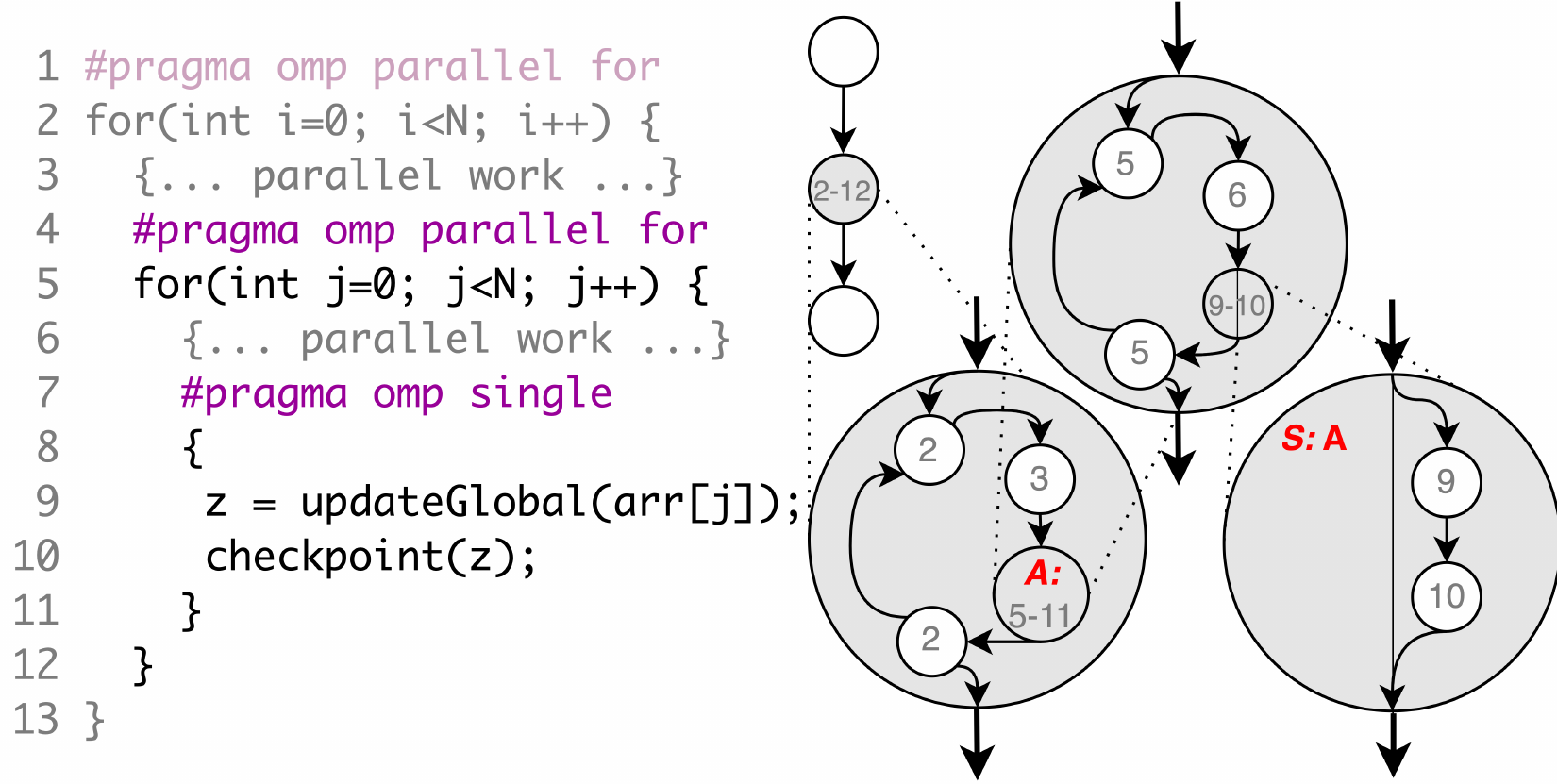}
\caption{Traits applied to context \emph{A} captures the region's semantics.}
\label{fig:def_context}
\end{figure}

\subsection{Context}
PPMs allow programmers to express semantics attached to a code region only when executed within the context of another code region.
For example, the code in a \code{single} OpenMP pragma needs to only be executed for one of the iterations of the innermost parallel loop that contains it.
It does not however specify that code should only be executed by a single iteration of an outer loop.
In other words, the parallel semantics of a \emph{single} section is valid only in the context of the innermost loop that contains it.
Because the contexts in which parallel semantics is valid cannot always be computed, the PS-PDG can specify contexts and their relation with parallel semantics.

A context in the PS-PDG represents a code region to which a parallel semantic applies.
A context in the PS-PDG is a labeled hierarchical node, where the label is a unique identifier.
Hence, hierarchical nodes of the PS-PDG that do not have a label are not contexts.
A parallel semantic explicitly lists the contexts in which it is valid, as shown in Tab.~\ref{tab:definition}. 
For example, the hierarchical node \emph{S} of Fig.~\ref{fig:def_context} that captures the \emph{single} code section declares its semantics applies only to context A, which is its target loop.

\subsection{Directed and Undirected Edges}
\begin{figure}[!h]
\centering
\includegraphics[keepaspectratio, width=.75\textwidth]{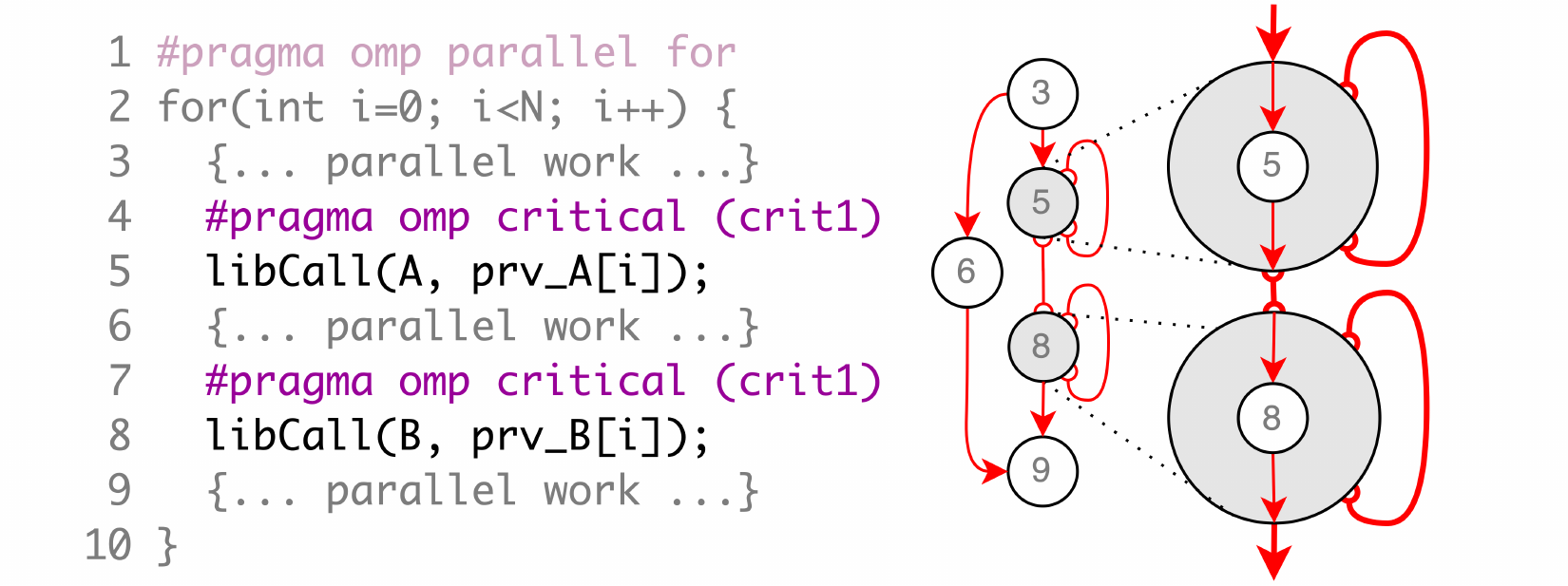}
\caption{Ordering constraints are captured with directed and undirected edges.}
\label{fig:def_edges}
\end{figure}

Parallel programming allows the declaration that two code regions (or two instructions) depend on each other but their relative execution order is not important.
This enables efficient parallel executions by avoiding unnecessary synchronizations.
PS-PDG includes both directed and undirected dependences (edges) to capture this semantics (Fig.~\ref{fig:def_edges}).

A directed edge in a PS-PDG follows the semantics of the PDG abstraction where the execution of the destination of that edge must wait for the edge's source execution.
Instead, an undirected edge expresses a dependence between two computations (e.g., instructions) that cannot run in parallel, but any ordering of their execution is allowed.

\subsection{Data-Selector Directed Edge}
\begin{figure}[!h]
\centering
\includegraphics[keepaspectratio, width=.75\textwidth]{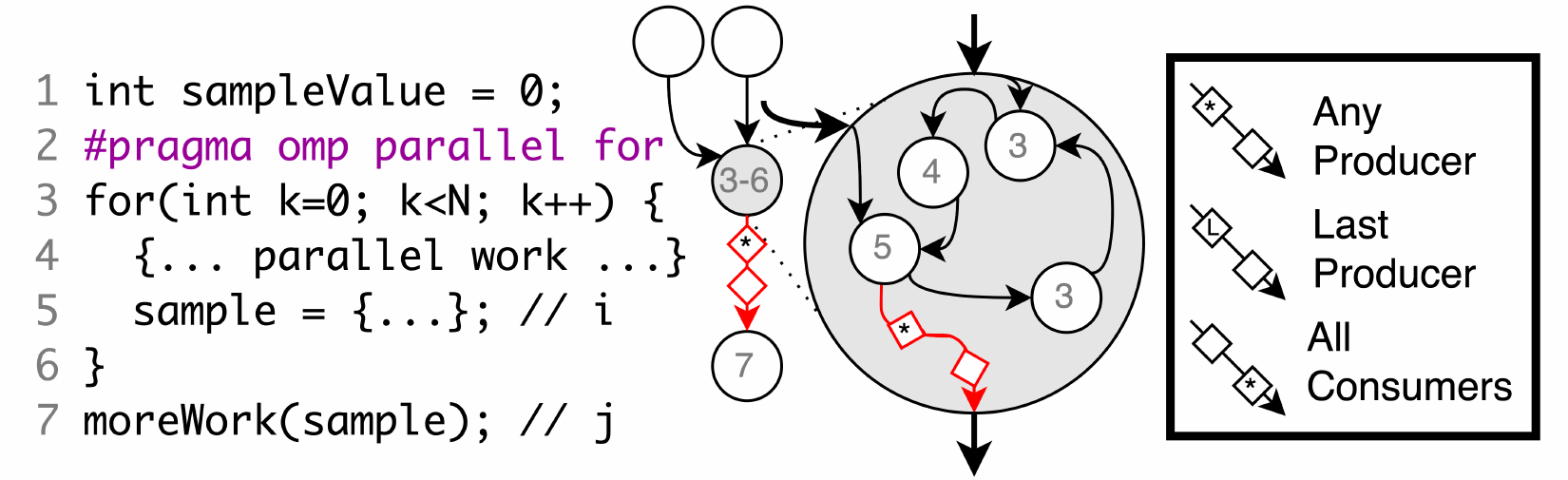}
\caption{Data-selector directed edges can capture non-trivial data relations.}
\label{fig:def_selector}
\end{figure}

The execution of an application typically includes many instances of a single static instruction (e.g., multiple executions of a single static instruction within a loop).
There is always a clear producer-consumer relation between dependent instructions for sequential programs.
For example, consider the instructions $i$ and $j$ shown in Fig.~\ref{fig:def_selector} which has a dependence from $i$ to $j$.
In this sequential program, the last instance of $i$ executed before $j$ will generate the data consumed by $j$ (this is captured by the PDG).
However, programmers can express richer semantics when developing a parallel program.
For example, a programmer can express that the data generated by any instance of $i$ can be used by $j$.
This is not expressible in prior abstractions like the PDG.
So, the PS-PDG introduces data-selectors that can be attached to a direct dependence.

A data-selector defines the set of dynamic instances of a static instruction.
A directed edge in the PS-PDG can have up to two data-selectors: one per static instruction attached to the edge.
A data-selector of the producer of a dependence defines which dynamic instance(s) of that producer are allowed to generate the data that will unlock the consumer.

This paper implements only the data-selectors required to capture the semantics of OpenMP and Cilk, which are the following:
\begin{itemize} 
    \item \emph{Any Producer Selector}: The consumer may use data generated by any instance of the producer. 
    \item \emph{Last Producer Selector}: The consumer must use data generated by the last instance of the producer.
    \item \emph{All Consumers Selector}: All consumers must use the data generated by the producer.
\end{itemize}

\subsection{Parallel Semantic Variables}
Efficient parallel execution often requires programmers to express knowledge about the program's variables that go beyond their reads and writes and their data types.
For example, programmers can express that a variable can be privatized in threads/tasks and all private instances can be merged (reduced) using application-specific knowledge.
This semantics goes beyond what can be expressed in sequential programming and therefore beyond what the PDG can capture (as the PDG was designed for sequential code).
To preserve this semantic, the PS-PDG introduces the concept of variables and their parallel semantics (how to clone them, their identity value, and how to reduce them) in its abstraction.

\begin{figure}[b]
\centering
\includegraphics[keepaspectratio, width=.75\textwidth]{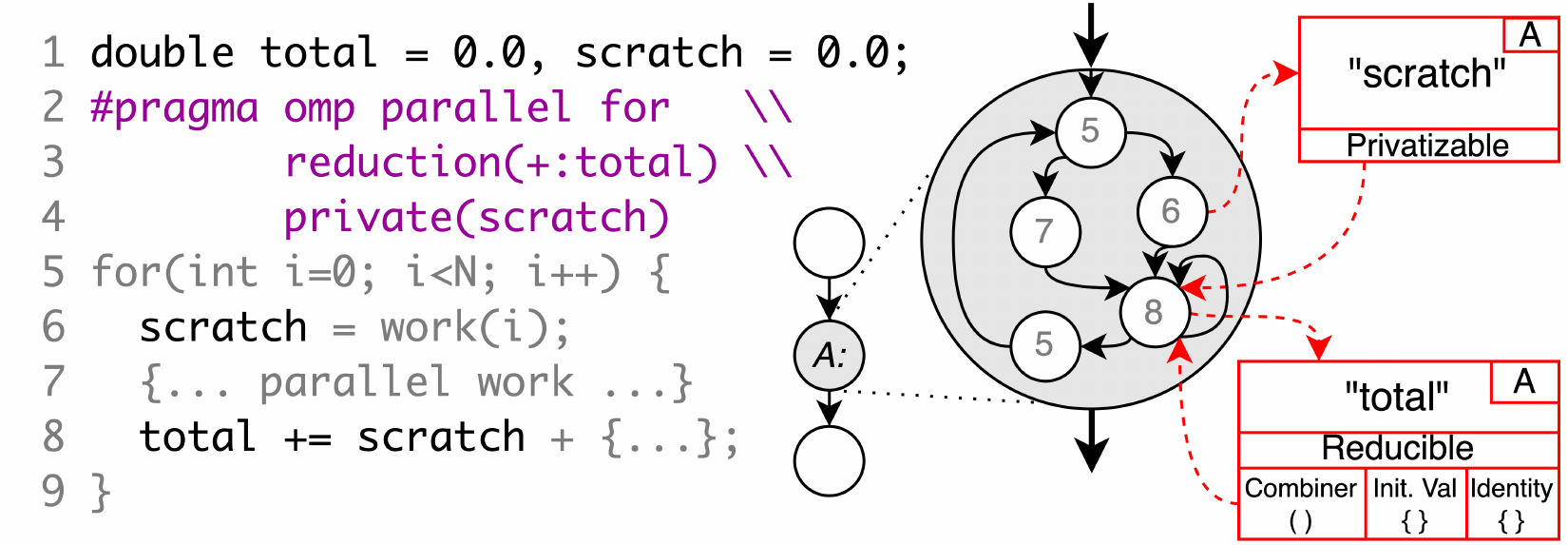}
\caption{Capturing programmer knowledge about data through privatizable and reducible parallel semantic variables.}
\label{fig:def_var_def_use}
\end{figure}

A parallel semantic variable in PS-PDG represents a variable or memory object that can be cloned to create private copies that a thread or task can independently use and modify.
This extension includes the code to execute to merge pairs of private copies together.
To do so, the variable description includes the reference to a computational node of the PS-PDG that represents a function.
This function takes two copies of a variable and it updates the first one with the result of the merge.
This merging operation is what compilers can use to reduce all private copies of a variable into a single one. 
An example of parallel semantic variable is shown in Fig.~\ref{fig:def_var_def_use}.

Parallel semantic variables are accessed by computation (e.g., an instruction).
Because such variables can be stored in memory, their accesses are not captured by the conventional use-def chains~\cite{aho2007compilers}.
To preserve this relation, the PS-PDG adds the Use/Def edges from a variable to PS-PDG nodes to encode the semantics that a target node uses and/or defines the variable at the source of that edge.

\section{The Necessity of Each PS-PDG Extension}
\label{sec:theoretical_evaluation}

This section demonstrates that each feature of the PS-PDG abstraction is necessary to capture the semantics expressible using the OpenMP programming model.
The same result can be obtained similarly for Cilk.
It does so for each PS-PDG extension by removing it from the proposed abstraction.
This is done by showing two parallel programs that have different parallelization plans and semantics.
By showing that these two programs translate to the same PS-PDG when the extension under evaluation is not available, we demonstrate that the feature is necessary.
Additionally, this section provides an example that shows how each feature enables an important optimization.  
This shows the value of each PS-PDG extension.

\begin{figure*}[hbtp]

\includegraphics[keepaspectratio,
width=\textwidth]
{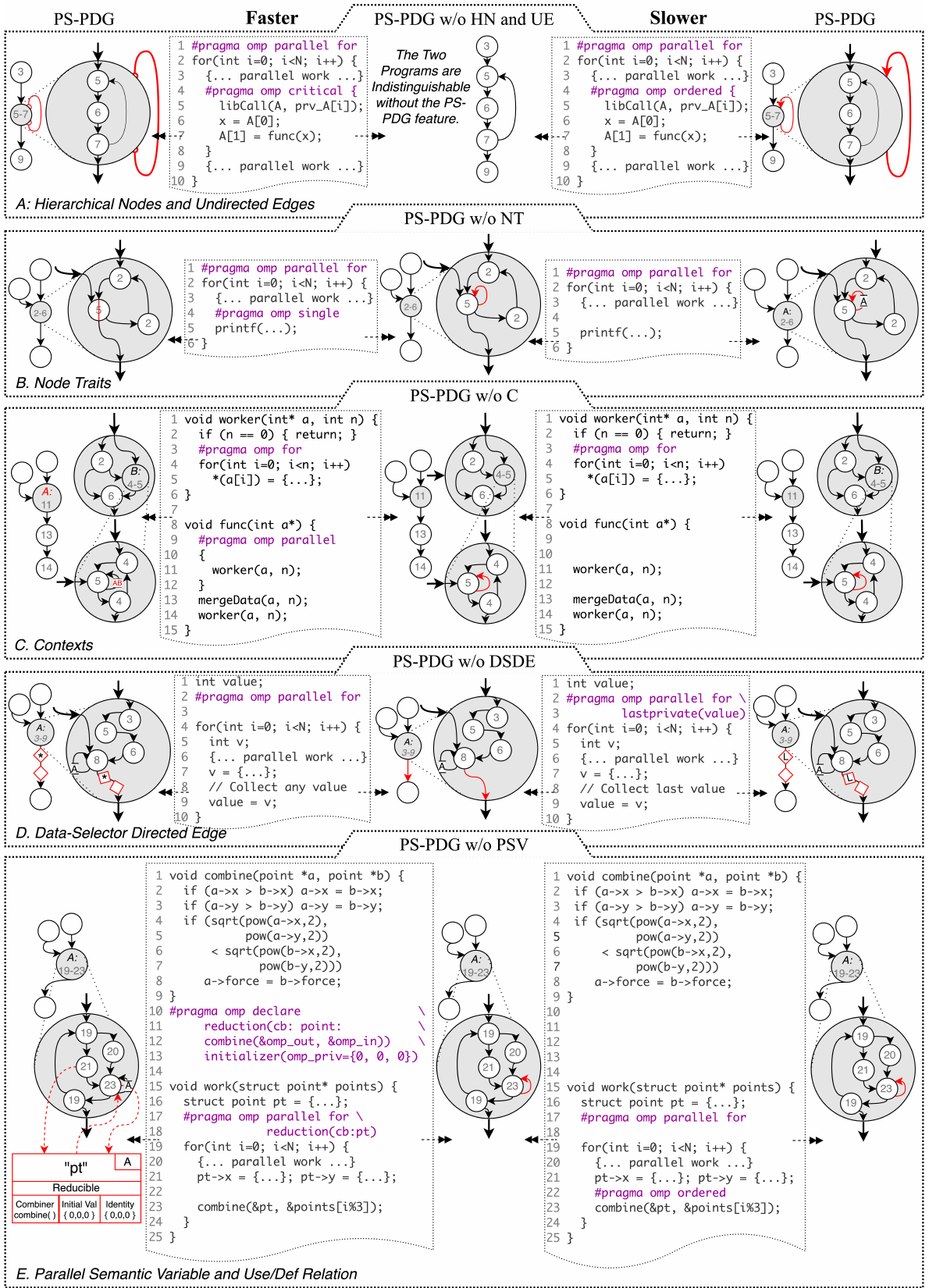}
\caption{Each feature of the PS-PDG is necessary since the removal of any PS-PDG feature would result in a loss of information. Without the given feature, the resulting abstraction is indistinguishable for the faster code (left) and the slower code (right).}
\label{fig:theory_eval_all_in_one}
\end{figure*}

\subsection{Hierarchical Nodes and Undirected Edges}

To understand the value of Hierarchical Nodes (HN) and Undirected Edges (UE), consider the two semantically different programs shown in Fig.~\ref{fig:theory_eval_all_in_one}-\emph{A}.
The program on the left requires avoiding overlapping dynamic instances of the critical section but puts no restriction on their order.
In contrast, the program on the right requires each dynamic instance of its critical section to be executed in loop-iteration order.
The program on the left executes significantly faster than the one on the right because it does not require synchronizations to enforce this additional constraint.
A compiler seeking the best parallelization plan for each program different in this one way must know whether or not this extra degree of freedom (orderless) exists.   In the PS-PDG, the undirected edge and hierarchical node features combined remove the ordering constraint while ensuring that dynamic instances of the connected nodes do not overlap.  When this feature is removed, this semantic information is lost.  Fig.~\ref{fig:theory_eval_all_in_one}-\emph{A} demonstrates this by showing how these two programs map to the same PS-PDG lacking these features (``PS-PDG w/o HN and UE").
Furthermore, this orderless semantics cannot be represented by the ``PS-PDG w/o HN and UE" because the orderless semantics does not hold at the single instruction granularity.

\subsection{Node Traits}
A node in the PS-PDG can hold various traits expressed
in a parallel programming language.  These traits can be important for the correctness
and performance of the parallel application. To understand the value of Node Traits (NT), consider the two semantically different programs shown in Fig.~\ref{fig:theory_eval_all_in_one}-\emph{B}.
The program on the left requires the singular execution of the print 
statement, allowing for quick and simple output from the parallel 
application. In contrast, the program on the right does not include a $single$ annotation for its print statement, meaning multiple calls to \code{printf}.
To maintain correctness, then the compiler must understand how the \code{printf} fits into the parallelization plan.   
Unfortunately, this is not possible when using the PS-PDG without Node Traits (``PS-PDG w/o NT").    Fig.~\ref{fig:theory_eval_all_in_one}-\emph{B} demonstrates this by showing how these two programs map to the same ``PS-PDG w/o NT".
In the ``PS-PDG w/o NT", the single execution semantic is lost.
Further, the single execution trait cannot be determined by compiler analysis from any other aspect of the ``PS-PDG w/o NT".

\subsection{Contexts}

To understand the value of Contexts (C), consider the two programs shown in Fig.~\ref{fig:theory_eval_all_in_one}-\emph{C}.  
The program on the left executes the first call to \code{worker} in parallel while the program on the right executes it sequentially.
By leveraging the parallelism in the hardware, the left program executes significantly faster than the program on the right. 
To generate the best parallel execution plan for the first \code{worker} call, the compiler needs to know in which context(s) the independent loop iteration semantic holds.   PS-PDG Contexts represent the contexts in which code region parallel semantics hold.
Without PS-PDG Contexts, the two programs map to the same ``PS-PDG w/o C".  Using only the  ``PS-PDG w/o C" in this example, the compiler cannot know when the loop iterations are independent of each other and must assume they are not.

\subsection{Data-Selector Directed Edge}
Data selectors can be added to the directed edges of the PS-PDG abstraction.
Data selectors define which dynamic instance (or instances) of the source node can generate the data that the destination node needs.
To understand the value of Data-Selector Directed Edge (DSDE), consider the two parallel programs shown in Fig.~\ref{fig:theory_eval_all_in_one}-\emph{D}.
Their semantics are different. The program on the right enforces that the value of the live-out variable \code{value} that can propagate outside the loop has to be the one generated during the last iteration of that loop.  The program on the left of Fig.~\ref{fig:theory_eval_all_in_one}-\emph{D} allows the propagation of the value generated by any loop iteration.  
The program on the left adds an extra degree of freedom.  With this freedom, a consumer of the live-out variable can start before the end of the loop execution, allowing for more overlapping computation.
Unfortunately, a PS-PDG without Data-Selector Directed Edges (``PS-PDG w/o DSDE") cannot distinguish these cases.  This can be seen as both programs in Fig.~\ref{fig:theory_eval_all_in_one}-\emph{D} map to the same ``PS-PDG w/o DSDE".
Thus, for correctness, the parallelization plan generated from the ``PS-PDG w/o DSDE" must enforce the stricter semantics of the program, the slower program on the right.
Note that the DSDE semantics cannot be inferred by a code analysis on the ``PS-PDG w/o DSDE".

\subsection{Parallel Semantic Variable and Use/Def Relation}
The PS-PDG includes Parallel Semantic Variables and Use/Def Relations (PSV) to represent a variable or object upon which the developer has encoded parallel semantics (e.g., how to reduce an object between tasks).  
These variables are connected to their computation (reads and writes) through Use/Def edges from parallel variables to nodes in the PS-PDG.
Consider the two parallel programs shown in Fig.~\ref{fig:theory_eval_all_in_one}-\emph{E}.
The program on the left runs all iterations of the loop in parallel without any synchronization between them.
Each thread operates on a private copy of the struct \code{pt}, then all private copies are reduced into a single one to be propagated to the code after the loop.  
This reduction is performed using application-specific knowledge.
In contrast, the program on the right of Fig.~\ref{fig:theory_eval_all_in_one}-\emph{E} has a single copy of the struct \code{pt} shared among all iterations of a loop.
Accesses (reads and writes) of this array are synchronized using an ordered section.
The program on the left executes significantly faster than the one on the right because it does not require any synchronization between the loop iterations running in parallel.
A compiler that needs to decide the parallelization plan to apply to the program on the left of Fig.~\ref{fig:theory_eval_all_in_one}-\emph{E}  needs to be aware of the ability to privatize and reduce the struct ~\code{pt} to generate the best parallelization plan.
Unfortunately, this is not possible when using a PS-PDG without Parallel Semantic Variables (``PS-PDG w/o PSV").  This becomes clear by observing that both programs in Fig.~\ref{fig:theory_eval_all_in_one}-\emph{E}  map to the same ``PS-PDG w/o PSV".  This means that the parallelization plan generated from the ``PS-PDG w/o PSV" must enforce the stricter semantics of the program on the right of Fig.~\ref{fig:theory_eval_all_in_one}-\emph{E}  where all array accesses are ordered.
Furthermore, notice that the lost application-specific knowledge about the reduction of the struct \code{pt} cannot be inferred from the ``PS-PDG w/o PSV".

\section{The Sufficiency of the PS-PDG for OpenMP}
\label{sec:sufficiency}

This section demonstrates that the PS-PDG abstraction is sufficient to capture the precise parallel constraints of the OpenMP programming model (Appendix~\ref{sec:appendix} demonstrates it similarly for the Cilk language). 
We target the OpenMP 5.0 specification~\cite{openmp50} with the exclusions of execution control, the target offload abstraction, runtime calls, and tooling support. 
We excluded these features because the goal is to enable compilers to select a parallel execution plan for general-purpose CPUs. So, we exclude features that only control the amount of parallelism to generate without adding semantics (e.g., deciding the number of threads to use for a given loop).

We group the parallel semantics expressible using the OpenMP 5.0 specification into three groups: declaration of independence, data properties, and computational ordering features. The OpenMP parallel semantics enabled by each group is mapped to a set of PS-PDG abstraction extensions. The parallel execution plan explicitly encoded by OpenMP programmer is within the parallelization plan space generated by a compiler using the PS-PDG. That is, the PS-PDG is sufficient to capture the precise parallel constraints of OpenMP 

\subsection{Declaration of Independence}

Declaring the independence between code regions has a significant impact on the parallel semantics of the OpenMP programming language. This type of parallel semantics is the most used in OpenMP programs (empirically confirmed by well-established benchmark suites). The most typical example of this semantics is \code{omp for}, which declares the independence between loop iterations of all the code within the loop body not included in a critical section. Another example is \code{task}, which groups computation into multiple tasks, and their dependences (or lack of) are explicitly declared. Other examples with similar semantics are \code{taskloop}, \code{sections}, \code{simd}, and \code{workshare}; they all declare the existence of parallelism between code regions. Finally, OpenMP provides clauses that programmers use to declare when the declaration of parallelism (or dependence) is valid (in which contexts). For example, the programmer can declare that two code regions are independent only when executed within the context of a loop, but not when executed within the context of outer loops. These clauses include \code{barrier}, \code{flush}, \code{taskwait}, and \code{depobj}. 
These clauses, such as \code{barrier}, do not add additional information, rather they constrain the information provided by other clauses.
This allows for the developer to encode synchronization in order to respect dependences which would instead be declared independent.
The PS-PDG captures these constraints as dependeces.

The semantics encoded by all declarations of independence of the OpenMP language is captured by the PS-PDG abstraction extensions \emph{hierarchical nodes} and \emph{contexts}. Each code region targeted by an OpenMP pragma mentioned above (e.g., \code{task}) is mapped into a hierarchical node of the PS-PDG abstraction. Edges between hierarchical nodes (including between a node and itself) declare their dependences as specified by the OpenMP programmer. Finally, dependence edges include contexts in the PS-PDG abstraction to declare when they are valid (and therefore when they are not). This captures the precise parallel constraints of all declarations of independence that programmers can do using OpenMP. 

For example, assume there is two-level nested loop where the inner loop has a call to a library function and that this call has a self-dependence only between iterations of the outer loop. An OpenMP programmer can parallelize this program by adding the pragma \code{for} to the inner loop. This semantics is mapped to PS-PDG creating a hierarchical node for the outer loop; hence, the outer loop becomes a context. Then, the dependence from the library call to itself includes the context of the outer loop, which declares that is valid only for the outer loop. A compiler can now generate the parallelization plan chosen by the OpenMP programmer using this PS-PDG as the inner loop has no loop-carried dependence and therefore can be parallelized using the parallelization plan selected by adding the pragma \code{for} to the inner loop.

\subsection{Data and its Properties}
The parallel semantics of data is essential for a PPM to be widely adopted.
Parallel semantics of data allows programmers to declare properties of data, how to use them, and how to propagate them throughout the computation.
An example of this parallel semantics is the \code{threadprivate} pragma, which declares that the data is attached to (e.g., an array) needs to be cloned such that threads use/define only their own private copy of it. 
Without access to privatization and reduction techniques 
the amount of expressible parallelism would be greatly limited.
Therefore, the OpenMP model (like other PPMs) enables programmers to declare how the data can be privatized per-thread (\code{threadprivate}), how can the private copies be reduced at the end of a parallel code region (\code{reduction}), and which data needs to be propagated throughout the computation (\code{first/last private}).

The OpenMP semantics about privatizing data and reducing their private copies is captured by the \emph{parallel semantics variable} of the PS-PDG abstraction. The PS-PDG variable declares its properties explicitly (e.g., per-thread private) and it declares how to reduce its private copies to a single one. The Use/Def relations included in PS-PDG declare how the code uses the related variable, and therefore when it needs to be privatized and when its private copies need to be reduced.

Finally, the OpenMP semantics about which data to propagate throughout the computation is captured by the \emph{data selectors} of the PS-PDG abstraction. These selectors declare which data needs to propagate from a producer to its consumers.

\subsection{Ordering}
OpenMP programmers can express that two code regions are dependent, but their execution order is not important.
This is more efficient than enforcing a pre-defined order.
This semantics is expressed using the \code{critical} or \code{atomic} pragmas.
The latter also imposes the need to execute atomically the code region wrapped in it. 
The PS-PDG \emph{undirected edge} declares this lack of ordering between two interdependent code regions (or instructions), including self-dependences, and the OpenMP atomic semantic is captured by the \emph{node trait} atomic.

\section{Evaluating the PS-PDG Abstraction}
\label{sec:system_methodology}
The PS-PDG abstraction captures the precise parallel constraints of a parallel program.
The strength of an abstraction used within a compiler is in its ability to represent knowledge about a program not readily ascertainable from the IR.
The value of a specific abstraction is in what it enables, and should be evaluated in this way.
To this end, we evaluate the PS-PDG by what it enables for an \emph{existing} automatic-parallelizing compiler rather than the end result of the parallelization.
Notice that evaluating on the end result of a transformation would evaluate the transformation implemented using the PS-PDG, rather than the expressiveness of the PS-PDG itself.
Hence, we evaluate the PS-PDG by performing two experiments that measure the power of the abstraction.
The first experiment measures the size of the PS-PDG enabled expansion of options for the parallelizing compiler.
The second experiment characterizes a bound on the potential of the parallel execution plans that the PS-PDG exposed.
We find the PS-PDG enabled compiler has a richer set of options when determining a parallel execution plan and exposes significantly better parallel execution plans.
Both experiments use a parallelizing compiler enhanced to use the PS-PDG in place of the PDG.
We utilize the entire \code{NAS} Benchmark Suite~\cite{nas} with class C inputs, with two exceptions (\code{BT,FT}: class B) due to gigabyte-size static variables.

\subsection{Implementing the PS-PDG in an existing compiler} \label{ssec:impl}
\begin{figure}[t]
\centering
\includegraphics[keepaspectratio, width=.99\textwidth]{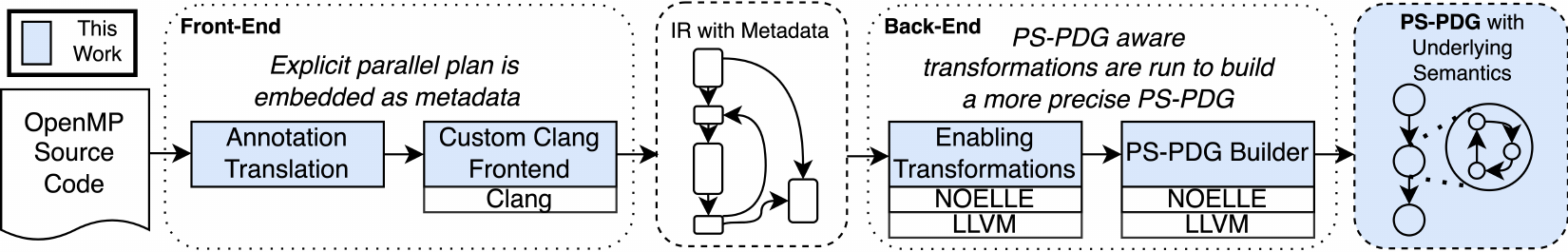}
\caption{Our custom compilation pipeline to generate the PS-PDG for a given OpenMP program.}
\label{fig:system_pipe}
\end{figure}

To perform the two experiments described above, we implemented a custom compilation pipeline (Fig.~\ref{fig:system_pipe}).  
This pipeline is built upon the NOELLE~\cite{matni2022noelle} compilation framework, which extends LLVM~\cite{llvm}.
NOELLE provides an automatic-parallelizing compiler working at the IR level, which supports three loop-based parallelization techniques: DOALL~\cite{doall}, DSWP~\cite{dswp}, HELIX~\cite{helix,helix2}.
First, we added a tool that translates OpenMP annotations into \code{pragmas} that are amenable to being lowered by our custom front-end.
Then, our custom clang-based front-end that generates LLVM IR with custom metadata from these pragmas.
This IR with metadata feeds a series of code transformations that are designed to make the code more amenable to parallelization while maintaining the metadata for the precise parallel constraints of the parallel program.
Finally, the resulting IR with metadata is used to build the PS-PDG.
The PS-PDG is then used by our extensions of NOELLE's compiler, which originally used the PDG.

Both our parallelizing compiler and NOELLE's original one consider the parallelization of each loop with at least $1\%$ run-time coverage.
The subset of a dependence graph (PS-PDG or PDG) for a given loop is analyzed to identify strongly-connected components (SCC) with loop-carried dependences.  
For these SCCs, we utilize any PS-PDG features within the SCC to determine if the loop-carried dependences can be removed (e.g., privatization). 
If a loop can be parallelized as DOALL (i.e., no loop-carried dependences with a known trip count), then it is only considered as DOALL. For non-DOALL loops, the compiler considers HELIX and DSWP.

\subsection{The PS-PDG gives the compiler more choices} \label{ssec:space}
To understand the potential added by PS-PDG, we automatically enumerate the options our compiler considers when determining a parallel execution plan of a loop.
Then, we compare it with those that the PDG-based compiler included in NOELLE has.
The PDG-based compiler utilizes the sequential version of the benchmarks. 
Additionally, we include the results from utilizing the OpenMP worksharing loop information for improved loop dependence analysis as in~\cite{jensen}, labeled as "J\&K".
This approach enables the worksharing loop information to remove loop-carried dependences in the PDG.
Finally, we include the number of corresponding options available to the source code OpenMP parallelization through environment variables.

We automatically enumerate the options for a 56 core machine while following the existing parallelization process in the compiler.
For DOALL loops, the number of options is at most 56 (cores) $\times$ 8 (chunk sizes considered). 
For non-DOALL loops, the SCCs of a loop are categorized as sequential or parallel.
The options available to HELIX is the possible number of sequential segments of that loop (a sequential segment is a slice of the loop that includes at least one sequential SCC).
Furthermore, we consider running these sequential segments in parallel up to 56 cores.
The options available to DSWP is the number of pipeline stages (each stage has at least one SCC) up to 56 cores.

\begin{figure}
\centering
\includegraphics[keepaspectratio, width=.95\textwidth]{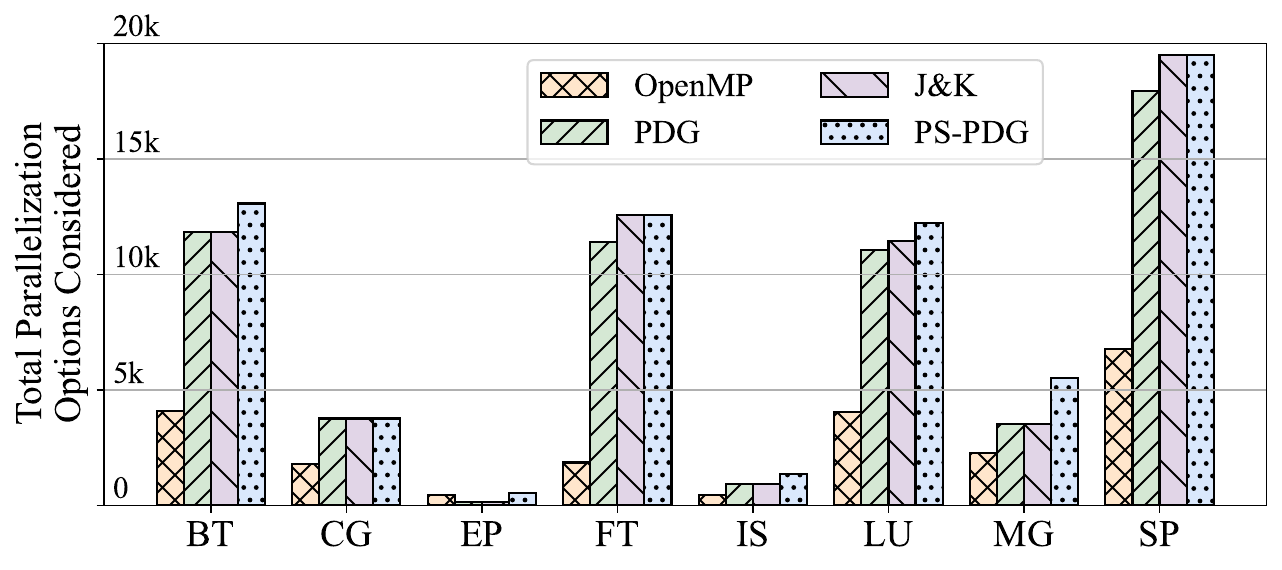}
\caption{Number of parallelization options available to the compiler.}
\label{fig:eval1}
\end{figure}

Fig.~\ref{fig:eval1} shows the number of options available to the compiler.
The PS-PDG enables the automatic-parallelizing compiler to explore more options than the PDG, resulting in a large increase in the number of parallelization plans when combined. 
The compiler is able to consider all loops which meet the parallelization requirements while the programmer-encoded parallelization is static, the PS-PDG pushes beyond the limitations of the PDG-based compiler.  
For benchmarks with few loops which are parallelized well by the programmer (e.g., \code{EP}), the increase in options stays low.
Additionally, we find that utilizing the PDG with workshare improved loop dependence analysis is insufficient to match the PS-PDG, as seen in the \code{MG} benchmark.
Through the PS-PDG, the compiler can leverage the precise parallel constraints of the explicit parallelism, while leveraging compiler analysis for a larger space of parallelization plans. 

\subsection{The parallelization potential exposed by the PS-PDG} \label{ssec:speedup}
Next we evaluate the quality of the additional parallelization plans these options create.
To evaluate the potential of the parallelization plans, we measure, via an emulator,  the critical path of the available parallelism on an ideal machine with unlimited cores, zero cost communication, and perfect memory access.
This enables us to characterize the limit of the additional parallelism unlocked by PS-PDG.
We do so by comparing the limit of the parallelism expressed by programmers, with the one obtainable using PDG, and with the parallelism obtainable using the new PS-PDG abstraction.
The critical path is computed as the number of dynamic LLVM instructions that must run sequentially given a parallelization plan.
The PDG measurements assume that every outermost loop is parallelized using DOALL, HELIX, or DSWP using the SCCs generated from the PDG.
The J\&K measurements assume that every outermost loop is parallelized using the SCCs from the PDG along with inner developer-expressed loops.
Lastly, the PS-PDG measurements assume that every outer loop is parallelized using the SCCs from the PS-PDG, as well as inner developer-expressed loops.
We only consider the hierarchical parallelism possible with existing parallelizing compilers or as expressed by the developer.
This methodology is consistent with that proposed by others~\cite{iiswc}.

\begin{figure}[ht]
\centering
\includegraphics[keepaspectratio, width=.95\textwidth]{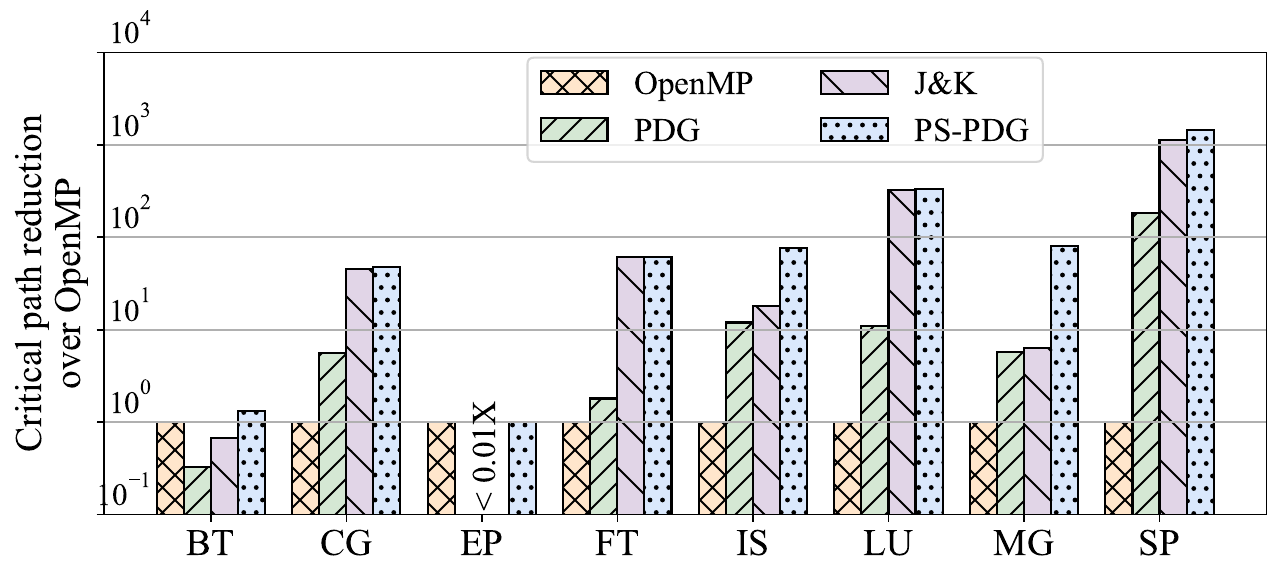}
\caption{Critical path reduction from abstraction-enabled parallelism.}
\label{fig:eval2}
\end{figure}

Fig.~\ref{fig:eval2} shows the critical path speedup over the programmer encoded parallelization for an ideal machine.
By utilizing the PS-PDG, the automatic-parallelizing compiler can leverage the developer encoded parallel semantics as well as the advanced compiler techniques, HELIX/DSWP.
The PS-PDG allows the compiler to discover plans with far more parallelization potential than with the PDG.
Additionally, for benchmarks with good parallelization coverage by the programmer (e.g., \code{EP}), the PS-PDG ensures no loss of parallelism since it captures the precise parallel constraints of the parallel program.
Finally, we find that utilizing workshare improved loop dependence analysis with the PDG (J\&K) is unable to unlock as much parallelization potential as the PS-PDG (e.g., \code{IS}). 

\section{Related Work}
\label{sec:related_work}
Previous work~\cite{sarkar_opt_ppg,ppg,wolfe_pcfg} proposed representations of the explicit parallelism encoded in a program to help programmers understand their parallelization.
These representations directly capture the parallel control flow encoded in the parallel program.
Other prior work~\cite{tapir,llvm-omp,inspire,psg,pst} lowered the explicit parallelism into the IR of the compiler, introducing a new IR where the parallel execution plan can be encoded explicitly.
Some prior work~\cite{jensen,DBLP:conf/sc/MaWNCHLHCHD21} analyzed parallel IRs that capture simple fork-join models to remove dependences from the PDG generated by compiler analyses (to unblock vectorization), but they do not handle semantics beyond simple fork-join.
Finally, HPVM~\cite{hpvm} is designed specifically for heterogeneous hardware to enable optimizations while still maintaining performance portability.
The PS-PDG is orthogonal to HPVM as it does not target heterogeneous hardware via a hierarchical dataflow graph or enable optimizations on the graph.

The Galois System~\cite{galois} and the Kinetic Dependence Graph~\cite{kdg} targeted implicit parallelism in imperative languages.
These approaches focus on irregular programs exploiting amorphous data-parallelism to improve performance by dynamically modifying computation task graphs at runtime.

Many functional programming languages represent parallelism either implicitly or explicitly through annotations or parallel constructs in the language itself (e.g., map)~\cite{blelloch_parallelism_1995,fluet_implicitly-threaded_nodate,westrick_disentanglement_2020,halstead_implementation_1984,arvind_i-structures_1989,blelloch_implementation_1993,li_lightweight_2007,marlow_parallel_2012,peyton_jones_harnessing_2008,guatto_hierarchical_2018,raghunathan_hierarchical_2016,sivaramakrishnan_multimlton_2014,Klusik_Loogen_Priebe_Rubio_2001}.
These works directly translated the parallelism into a single or a few predetermined parallel execution plans (usually based on task or fork-join parallelism), where the runtime system is left with few decisions to make (e.g., number of threads).

\section{Conclusion}
\label{sec:conclusion}
This work presents the PS-PDG, a novel abstraction that captures the precise parallel constraints of modern parallel programs.
The PS-PDG is shown to be necessary and sufficient for capturing the precise parallel constraints found in OpenMP and Cilk.
Our PS-PDG enabled compiler is capable of enabling advanced parallel execution plan optimizations for explicitly parallel programs on modern many-core systems.

\nocite{rtl}
\clearpage
\bibliographystyle{ACM-Reference-Format}
\bibliography{references.bib}


\begin{thebibliography}{57}


\ifx \showCODEN    \undefined \def \showCODEN     #1{\unskip}     \fi
\ifx \showDOI      \undefined \def \showDOI       #1{#1}\fi
\ifx \showISBNx    \undefined \def \showISBNx     #1{\unskip}     \fi
\ifx \showISBNxiii \undefined \def \showISBNxiii  #1{\unskip}     \fi
\ifx \showISSN     \undefined \def \showISSN      #1{\unskip}     \fi
\ifx \showLCCN     \undefined \def \showLCCN      #1{\unskip}     \fi
\ifx \shownote     \undefined \def \shownote      #1{#1}          \fi
\ifx \showarticletitle \undefined \def \showarticletitle #1{#1}   \fi
\ifx \showURL      \undefined \def \showURL       {\relax}        \fi
\providecommand\bibfield[2]{#2}
\providecommand\bibinfo[2]{#2}
\providecommand\natexlab[1]{#1}
\providecommand\showeprint[2][]{arXiv:#2}

\bibitem[ope(2022)]%
        {opencilk}
 \bibinfo{year}{2022}\natexlab{}.
\newblock \bibinfo{title}{Write fast code with C/C++ and OpenCilk.}
\newblock
\newblock
\urldef\tempurl%
\url{https://www.opencilk.org/}
\showURL{%
\tempurl}


\bibitem[llv(2023)]%
        {llvm-omp}
 \bibinfo{year}{2023}\natexlab{}.
\newblock \bibinfo{title}{{LLVM/OpenMP} design and overview}.
\newblock
\newblock
\urldef\tempurl%
\url{https://www.https://openmp.llvm.org/}
\showURL{%
\tempurl}


\bibitem[rtl(2023)]%
        {rtl}
 \bibinfo{year}{2023}\natexlab{}.
\newblock \bibinfo{title}{{RTL} Representation}.
\newblock
\newblock
\urldef\tempurl%
\url{https://gcc.gnu.org/onlinedocs/gccint/RTL.html}
\showURL{%
\tempurl}


\bibitem[Aho et~al\mbox{.}(2007)]%
        {aho2007compilers}
\bibfield{author}{\bibinfo{person}{Alfred~V Aho}, \bibinfo{person}{Monica~S
  Lam}, \bibinfo{person}{Ravi Sethi}, {and} \bibinfo{person}{Jeffrey~D
  Ullman}.} \bibinfo{year}{2007}\natexlab{}.
\newblock \bibinfo{booktitle}{\emph{Compilers: principles, techniques, \&
  tools}}.
\newblock \bibinfo{publisher}{Pearson Education India}.
\newblock


\bibitem[Apostolakis et~al\mbox{.}(2020a)]%
        {perspective}
\bibfield{author}{\bibinfo{person}{Sotiris Apostolakis},
  \bibinfo{person}{Ziyang Xu}, \bibinfo{person}{Greg Chan},
  \bibinfo{person}{Simone Campanoni}, {and} \bibinfo{person}{David~I. August}.}
  \bibinfo{year}{2020}\natexlab{a}.
\newblock \showarticletitle{Perspective: A Sensible Approach to Speculative
  Automatic Parallelization}. In \bibinfo{booktitle}{\emph{Proceedings of the
  Twenty-Fifth International Conference on Architectural Support for
  Programming Languages and Operating Systems}} (Lausanne, Switzerland)
  \emph{(\bibinfo{series}{ASPLOS '20})}. \bibinfo{publisher}{Association for
  Computing Machinery}, \bibinfo{address}{New York, NY, USA},
  \bibinfo{pages}{351–367}.
\newblock
\showISBNx{9781450371025}
\urldef\tempurl%
\url{https://doi.org/10.1145/3373376.3378458}
\showDOI{\tempurl}


\bibitem[Apostolakis et~al\mbox{.}(2020b)]%
        {scaf}
\bibfield{author}{\bibinfo{person}{Sotiris Apostolakis},
  \bibinfo{person}{Ziyang Xu}, \bibinfo{person}{Zujun Tan},
  \bibinfo{person}{Greg Chan}, \bibinfo{person}{Simone Campanoni}, {and}
  \bibinfo{person}{David~I. August}.} \bibinfo{year}{2020}\natexlab{b}.
\newblock \showarticletitle{{SCAF:} A Speculation-Aware Collaborative
  Dependence Analysis Framework}. In \bibinfo{booktitle}{\emph{Proceedings of
  the 41st ACM SIGPLAN Conference on Programming Language Design and
  Implementation}} (London, UK) \emph{(\bibinfo{series}{PLDI 2020})}.
  \bibinfo{publisher}{Association for Computing Machinery},
  \bibinfo{address}{New York, NY, USA}, \bibinfo{pages}{638–654}.
\newblock
\showISBNx{9781450376136}
\urldef\tempurl%
\url{https://doi.org/10.1145/3385412.3386028}
\showDOI{\tempurl}


\bibitem[Arvind et~al\mbox{.}(1989)]%
        {arvind_i-structures_1989}
\bibfield{author}{\bibinfo{person}{Arvind Arvind}, \bibinfo{person}{Rishiyur
  Nikhil}, {and} \bibinfo{person}{Keshav Pingali}.}
  \bibinfo{year}{1989}\natexlab{}.
\newblock \showarticletitle{I-{Structures}: Data Structures for Parallel
  Computing.}
\newblock \bibinfo{journal}{\emph{ACM Trans. Program. Lang. Syst.}}
  \bibinfo{volume}{11} (\bibinfo{date}{Oct.} \bibinfo{year}{1989}),
  \bibinfo{pages}{598--632}.
\newblock
\urldef\tempurl%
\url{https://doi.org/10.1145/69558.69562}
\showDOI{\tempurl}


\bibitem[Bailey et~al\mbox{.}(1991)]%
        {nas}
\bibfield{author}{\bibinfo{person}{D.~H. Bailey}, \bibinfo{person}{E. Barszcz},
  \bibinfo{person}{J.~T. Barton}, \bibinfo{person}{D.~S. Browning},
  \bibinfo{person}{R.~L. Carter}, \bibinfo{person}{L. Dagum},
  \bibinfo{person}{R.~A. Fatoohi}, \bibinfo{person}{P.~O. Frederickson},
  \bibinfo{person}{T.~A. Lasinski}, \bibinfo{person}{R.~S. Schreiber},
  \bibinfo{person}{H.~D. Simon}, \bibinfo{person}{V. Venkatakrishnan}, {and}
  \bibinfo{person}{S.~K. Weeratunga}.} \bibinfo{year}{1991}\natexlab{}.
\newblock \showarticletitle{The {NAS} Parallel Benchmarks—Summary and
  Preliminary Results}. In \bibinfo{booktitle}{\emph{Proceedings of the 1991
  ACM/IEEE Conference on Supercomputing}} (Albuquerque, New Mexico, USA)
  \emph{(\bibinfo{series}{Supercomputing '91})}.
  \bibinfo{publisher}{Association for Computing Machinery},
  \bibinfo{address}{New York, NY, USA}, \bibinfo{pages}{158–165}.
\newblock
\showISBNx{0897914597}
\urldef\tempurl%
\url{https://doi.org/10.1145/125826.125925}
\showDOI{\tempurl}


\bibitem[Beckingsale et~al\mbox{.}(2019)]%
        {raja}
\bibfield{author}{\bibinfo{person}{David~A. Beckingsale},
  \bibinfo{person}{Jason Burmark}, \bibinfo{person}{Rich Hornung},
  \bibinfo{person}{Holger Jones}, \bibinfo{person}{William Killian},
  \bibinfo{person}{Adam~J. Kunen}, \bibinfo{person}{Olga Pearce},
  \bibinfo{person}{Peter Robinson}, \bibinfo{person}{Brian~S. Ryujin}, {and}
  \bibinfo{person}{Thomas~RW Scogland}.} \bibinfo{year}{2019}\natexlab{}.
\newblock \showarticletitle{RAJA: Portable Performance for Large-Scale
  Scientific Applications}. In \bibinfo{booktitle}{\emph{2019 IEEE/ACM
  International Workshop on Performance, Portability and Productivity in HPC
  (P3HPC)}}. \bibinfo{pages}{71--81}.
\newblock
\urldef\tempurl%
\url{https://doi.org/10.1109/P3HPC49587.2019.00012}
\showDOI{\tempurl}


\bibitem[Blelloch and Greiner(1995)]%
        {blelloch_parallelism_1995}
\bibfield{author}{\bibinfo{person}{Guy Blelloch} {and} \bibinfo{person}{John
  Greiner}.} \bibinfo{year}{1995}\natexlab{}.
\newblock \showarticletitle{Parallelism in sequential functional languages}. In
  \bibinfo{booktitle}{\emph{Proceedings of the seventh international conference
  on {Functional} programming languages and computer architecture - {FPCA}
  '95}}. \bibinfo{publisher}{ACM Press}, \bibinfo{address}{La Jolla,
  California, United States}, \bibinfo{pages}{226--237}.
\newblock
\showISBNx{978-0-89791-719-3}
\urldef\tempurl%
\url{https://doi.org/10.1145/224164.224210}
\showDOI{\tempurl}


\bibitem[Blelloch et~al\mbox{.}(1993)]%
        {blelloch_implementation_1993}
\bibfield{author}{\bibinfo{person}{Guy~E. Blelloch},
  \bibinfo{person}{Jonathan~C. Hardwick}, \bibinfo{person}{Siddhartha
  Chatterjee}, \bibinfo{person}{Jay Sipelstein}, {and} \bibinfo{person}{Marco
  Zagha}.} \bibinfo{year}{1993}\natexlab{}.
\newblock \showarticletitle{Implementation of a portable nested data-parallel
  language}.
\newblock \bibinfo{journal}{\emph{ACM SIGPLAN Notices}} \bibinfo{volume}{28},
  \bibinfo{number}{7} (\bibinfo{date}{July} \bibinfo{year}{1993}),
  \bibinfo{pages}{102--111}.
\newblock
\showISSN{0362-1340}
\urldef\tempurl%
\url{https://doi.org/10.1145/173284.155343}
\showDOI{\tempurl}


\bibitem[Blumofe et~al\mbox{.}(1996)]%
        {cilk}
\bibfield{author}{\bibinfo{person}{Robert~D Blumofe},
  \bibinfo{person}{Christopher~F Joerg}, \bibinfo{person}{Bradley~C Kuszmaul},
  \bibinfo{person}{Charles~E Leiserson}, \bibinfo{person}{Keith~H Randall},
  {and} \bibinfo{person}{Yuli Zhou}.} \bibinfo{year}{1996}\natexlab{}.
\newblock \showarticletitle{Cilk: An efficient multithreaded runtime system}.
\newblock \bibinfo{journal}{\emph{Journal of parallel and distributed
  computing}} \bibinfo{volume}{37}, \bibinfo{number}{1} (\bibinfo{year}{1996}),
  \bibinfo{pages}{55--69}.
\newblock


\bibitem[Campanoni and Crespi~Reghizzi(2009)]%
        {10.1007/978-3-642-02737-6_12}
\bibfield{author}{\bibinfo{person}{Simone Campanoni} {and}
  \bibinfo{person}{Stefano Crespi~Reghizzi}.} \bibinfo{year}{2009}\natexlab{}.
\newblock \showarticletitle{Traces of Control-Flow Graphs}. In
  \bibinfo{booktitle}{\emph{Developments in Language Theory}},
  \bibfield{editor}{\bibinfo{person}{Volker Diekert} {and}
  \bibinfo{person}{Dirk Nowotka}} (Eds.). \bibinfo{publisher}{Springer Berlin
  Heidelberg}, \bibinfo{address}{Berlin, Heidelberg},
  \bibinfo{pages}{156--169}.
\newblock
\showISBNx{978-3-642-02737-6}


\bibitem[Campanoni et~al\mbox{.}(2012a)]%
        {helix}
\bibfield{author}{\bibinfo{person}{Simone Campanoni}, \bibinfo{person}{Timothy
  Jones}, \bibinfo{person}{Glenn Holloway}, \bibinfo{person}{Vijay~Janapa
  Reddi}, \bibinfo{person}{Gu-Yeon Wei}, {and} \bibinfo{person}{David Brooks}.}
  \bibinfo{year}{2012}\natexlab{a}.
\newblock \showarticletitle{{HELIX}: Automatic Parallelization of Irregular
  Programs for Chip Multiprocessing}. In \bibinfo{booktitle}{\emph{Proceedings
  of the Tenth International Symposium on Code Generation and Optimization}}
  (San Jose, California) \emph{(\bibinfo{series}{CGO '12})}.
  \bibinfo{publisher}{ACM}, \bibinfo{address}{New York, NY, USA},
  \bibinfo{pages}{84--93}.
\newblock
\showISBNx{978-1-4503-1206-6}
\urldef\tempurl%
\url{https://doi.org/10.1145/2259016.2259028}
\showDOI{\tempurl}


\bibitem[Campanoni et~al\mbox{.}(2012b)]%
        {helix2}
\bibfield{author}{\bibinfo{person}{Simone Campanoni},
  \bibinfo{person}{Timothy~M Jones}, \bibinfo{person}{Glenn Holloway},
  \bibinfo{person}{Gu-Yeon Wei}, {and} \bibinfo{person}{David Brooks}.}
  \bibinfo{year}{2012}\natexlab{b}.
\newblock \showarticletitle{{HELIX:} Making the extraction of thread-level
  parallelism mainstream}.
\newblock \bibinfo{journal}{\emph{IEEE Micro}} \bibinfo{volume}{32},
  \bibinfo{number}{4} (\bibinfo{year}{2012}), \bibinfo{pages}{8--18}.
\newblock


\bibitem[Chang et~al\mbox{.}(2016)]%
        {pp_kernel_synth}
\bibfield{author}{\bibinfo{person}{Li-Wen Chang}, \bibinfo{person}{Izzat~El
  Hajj}, \bibinfo{person}{Christopher Rodrigues}, \bibinfo{person}{Juan
  Gómez-Luna}, {and} \bibinfo{person}{Wen-mei Hwu}.}
  \bibinfo{year}{2016}\natexlab{}.
\newblock \showarticletitle{Efficient kernel synthesis for performance portable
  programming}. In \bibinfo{booktitle}{\emph{2016 49th Annual IEEE/ACM
  International Symposium on Microarchitecture (MICRO)}}.
  \bibinfo{pages}{1--13}.
\newblock
\urldef\tempurl%
\url{https://doi.org/10.1109/MICRO.2016.7783715}
\showDOI{\tempurl}


\bibitem[Deakin et~al\mbox{.}(2019)]%
        {pp_diverse_arch}
\bibfield{author}{\bibinfo{person}{Tom Deakin}, \bibinfo{person}{Simon
  McIntosh-Smith}, \bibinfo{person}{James Price}, \bibinfo{person}{Andrei
  Poenaru}, \bibinfo{person}{Patrick Atkinson}, \bibinfo{person}{Codrin Popa},
  {and} \bibinfo{person}{Justin Salmon}.} \bibinfo{year}{2019}\natexlab{}.
\newblock \showarticletitle{Performance Portability across Diverse Computer
  Architectures}. In \bibinfo{booktitle}{\emph{2019 IEEE/ACM International
  Workshop on Performance, Portability and Productivity in HPC (P3HPC)}}.
  \bibinfo{pages}{1--13}.
\newblock
\urldef\tempurl%
\url{https://doi.org/10.1109/P3HPC49587.2019.00006}
\showDOI{\tempurl}


\bibitem[Fan et~al\mbox{.}(2019)]%
        {DBLP:conf/isca/FanCJ19}
\bibfield{author}{\bibinfo{person}{Yuanbo Fan}, \bibinfo{person}{Simone
  Campanoni}, {and} \bibinfo{person}{Russ Joseph}.}
  \bibinfo{year}{2019}\natexlab{}.
\newblock \showarticletitle{Time squeezing for tiny devices}. In
  \bibinfo{booktitle}{\emph{Proceedings of the 46th International Symposium on
  Computer Architecture, {ISCA} 2019, Phoenix, AZ, USA, June 22-26, 2019}}.
  \bibinfo{pages}{657--670}.
\newblock
\urldef\tempurl%
\url{https://doi.org/10.1145/3307650.3322268}
\showDOI{\tempurl}


\bibitem[Fan et~al\mbox{.}(2018)]%
        {Fan:2018:CIC:3195970.3196013}
\bibfield{author}{\bibinfo{person}{Yuanbo Fan}, \bibinfo{person}{Tianyu Jia},
  \bibinfo{person}{Jie Gu}, \bibinfo{person}{Simone Campanoni}, {and}
  \bibinfo{person}{Russ Joseph}.} \bibinfo{year}{2018}\natexlab{}.
\newblock \showarticletitle{Compiler-guided Instruction-level Clock Scheduling
  for Timing Speculative Processors}. In \bibinfo{booktitle}{\emph{Proceedings
  of the 55th Annual Design Automation Conference}} (San Francisco, California)
  \emph{(\bibinfo{series}{DAC '18})}. \bibinfo{publisher}{ACM},
  \bibinfo{address}{New York, NY, USA}, Article \bibinfo{articleno}{40},
  \bibinfo{numpages}{6}~pages.
\newblock
\showISBNx{978-1-4503-5700-5}
\urldef\tempurl%
\url{https://doi.org/10.1145/3195970.3196013}
\showDOI{\tempurl}


\bibitem[Ferrante et~al\mbox{.}(1987)]%
        {pdg}
\bibfield{author}{\bibinfo{person}{Jeanne Ferrante}, \bibinfo{person}{Karl~J.
  Ottenstein}, {and} \bibinfo{person}{Joe~D. Warren}.}
  \bibinfo{year}{1987}\natexlab{}.
\newblock \showarticletitle{The Program Dependence Graph and Its Use in
  Optimization}.
\newblock \bibinfo{journal}{\emph{ACM Trans. Program. Lang. Syst.}}
  \bibinfo{volume}{9}, \bibinfo{number}{3} (\bibinfo{date}{jul}
  \bibinfo{year}{1987}), \bibinfo{pages}{319–349}.
\newblock
\showISSN{0164-0925}
\urldef\tempurl%
\url{https://doi.org/10.1145/24039.24041}
\showDOI{\tempurl}


\bibitem[Fluet et~al\mbox{.}({[n.\,d.]})]%
        {fluet_implicitly-threaded_nodate}
\bibfield{author}{\bibinfo{person}{Matthew Fluet}, \bibinfo{person}{Mike
  Rainey}, \bibinfo{person}{John Reppy}, {and} \bibinfo{person}{Adam Shaw}.}
  \bibinfo{year}{[n.\,d.]}\natexlab{}.
\newblock \showarticletitle{Implicitly-threaded Parallelism in {Manticore}}.
\newblock  (\bibinfo{year}{[n.\,d.]}), \bibinfo{pages}{12}.
\newblock


\bibitem[Garcia et~al\mbox{.}(2011)]%
        {kremlin}
\bibfield{author}{\bibinfo{person}{Saturnino Garcia}, \bibinfo{person}{Donghwan
  Jeon}, \bibinfo{person}{Christopher~M. Louie}, {and}
  \bibinfo{person}{Michael~Bedford Taylor}.} \bibinfo{year}{2011}\natexlab{}.
\newblock \showarticletitle{Kremlin: Rethinking and Rebooting Gprof for the
  Multicore Age}. In \bibinfo{booktitle}{\emph{Proceedings of the 32nd ACM
  SIGPLAN Conference on Programming Language Design and Implementation}} (San
  Jose, California, USA) \emph{(\bibinfo{series}{PLDI '11})}.
  \bibinfo{publisher}{Association for Computing Machinery},
  \bibinfo{address}{New York, NY, USA}, \bibinfo{pages}{458–469}.
\newblock
\showISBNx{9781450306638}
\urldef\tempurl%
\url{https://doi.org/10.1145/1993498.1993553}
\showDOI{\tempurl}


\bibitem[Guatto et~al\mbox{.}(2018)]%
        {guatto_hierarchical_2018}
\bibfield{author}{\bibinfo{person}{Adrien Guatto}, \bibinfo{person}{Sam
  Westrick}, \bibinfo{person}{Ram Raghunathan}, \bibinfo{person}{Umut Acar},
  {and} \bibinfo{person}{Matthew Fluet}.} \bibinfo{year}{2018}\natexlab{}.
\newblock \showarticletitle{Hierarchical memory management for mutable state}.
\newblock \bibinfo{journal}{\emph{ACM SIGPLAN Notices}} \bibinfo{volume}{53},
  \bibinfo{number}{1} (\bibinfo{date}{Feb.} \bibinfo{year}{2018}),
  \bibinfo{pages}{81--93}.
\newblock
\showISSN{0362-1340}
\urldef\tempurl%
\url{https://doi.org/10.1145/3200691.3178494}
\showDOI{\tempurl}


\bibitem[Halstead(1984)]%
        {halstead_implementation_1984}
\bibfield{author}{\bibinfo{person}{Robert~H. Halstead}.}
  \bibinfo{year}{1984}\natexlab{}.
\newblock \showarticletitle{Implementation of multilisp: {Lisp} on a
  multiprocessor}. In \bibinfo{booktitle}{\emph{Proceedings of the 1984 {ACM}
  {Symposium} on {LISP} and functional programming}}
  \emph{(\bibinfo{series}{{LFP} '84})}. \bibinfo{publisher}{Association for
  Computing Machinery}, \bibinfo{address}{New York, NY, USA},
  \bibinfo{pages}{9--17}.
\newblock
\showISBNx{978-0-89791-142-9}
\urldef\tempurl%
\url{https://doi.org/10.1145/800055.802017}
\showDOI{\tempurl}


\bibitem[Hassaan et~al\mbox{.}(2015)]%
        {kdg}
\bibfield{author}{\bibinfo{person}{Muhammad~Amber Hassaan},
  \bibinfo{person}{Donald~D. Nguyen}, {and} \bibinfo{person}{Keshav~K.
  Pingali}.} \bibinfo{year}{2015}\natexlab{}.
\newblock \showarticletitle{Kinetic Dependence Graphs}. In
  \bibinfo{booktitle}{\emph{Proceedings of the Twentieth International
  Conference on Architectural Support for Programming Languages and Operating
  Systems}} (Istanbul, Turkey) \emph{(\bibinfo{series}{ASPLOS '15})}.
  \bibinfo{publisher}{Association for Computing Machinery},
  \bibinfo{address}{New York, NY, USA}, \bibinfo{pages}{457–471}.
\newblock
\showISBNx{9781450328357}
\urldef\tempurl%
\url{https://doi.org/10.1145/2694344.2694363}
\showDOI{\tempurl}


\bibitem[Hurson et~al\mbox{.}(1997)]%
        {doall}
\bibfield{author}{\bibinfo{person}{Ali~R Hurson}, \bibinfo{person}{Joford~T
  Lim}, \bibinfo{person}{Krishna~M Kavi}, {and} \bibinfo{person}{Ben Lee}.}
  \bibinfo{year}{1997}\natexlab{}.
\newblock \showarticletitle{Parallelization of {DOALL} and {DOACROSS} loops—a
  survey}.
\newblock In \bibinfo{booktitle}{\emph{Advances in computers}}.
  Vol.~\bibinfo{volume}{45}. \bibinfo{publisher}{Elsevier},
  \bibinfo{pages}{53--103}.
\newblock


\bibitem[ISO JTC1/SC22/WG14 - N1665(2012)]%
        {cilkplus_spec}
ISO JTC1/SC22/WG14 - N1665 \bibinfo{year}{2012}\natexlab{}.
\newblock \bibinfo{booktitle}{\emph{Intel® Cilk™ Plus Language Extension
  Specification}}.
\newblock \bibinfo{type}{{T}echnical {R}eport}.
  \bibinfo{institution}{International Organization for Standardization}.
\newblock
\urldef\tempurl%
\url{https://www.open-std.org/jtc1/sc22/wg14/www/docs/n1665.htm}
\showURL{%
\tempurl}


\bibitem[Jensen and Karlsson(2017)]%
        {jensen}
\bibfield{author}{\bibinfo{person}{Nicklas~Bo Jensen} {and}
  \bibinfo{person}{Sven Karlsson}.} \bibinfo{year}{2017}\natexlab{}.
\newblock \showarticletitle{Improving Loop Dependence Analysis}.
\newblock \bibinfo{journal}{\emph{ACM Trans. Archit. Code Optim.}}
  \bibinfo{volume}{14}, \bibinfo{number}{3}, Article \bibinfo{articleno}{22}
  (\bibinfo{date}{aug} \bibinfo{year}{2017}), \bibinfo{numpages}{24}~pages.
\newblock
\showISSN{1544-3566}
\urldef\tempurl%
\url{https://doi.org/10.1145/3095754}
\showDOI{\tempurl}


\bibitem[Johnson et~al\mbox{.}(1994)]%
        {pst}
\bibfield{author}{\bibinfo{person}{Richard Johnson}, \bibinfo{person}{David
  Pearson}, {and} \bibinfo{person}{Keshav Pingali}.}
  \bibinfo{year}{1994}\natexlab{}.
\newblock \showarticletitle{The Program Structure Tree: Computing Control
  Regions in Linear Time}.
\newblock \bibinfo{journal}{\emph{SIGPLAN Not.}} \bibinfo{volume}{29},
  \bibinfo{number}{6} (\bibinfo{date}{jun} \bibinfo{year}{1994}),
  \bibinfo{pages}{171–185}.
\newblock
\showISSN{0362-1340}
\urldef\tempurl%
\url{https://doi.org/10.1145/773473.178258}
\showDOI{\tempurl}


\bibitem[Jordan et~al\mbox{.}(2013)]%
        {inspire}
\bibfield{author}{\bibinfo{person}{Herbert Jordan}, \bibinfo{person}{Simone
  Pellegrini}, \bibinfo{person}{Peter Thoman}, \bibinfo{person}{Klaus Kofler},
  {and} \bibinfo{person}{Thomas Fahringer}.} \bibinfo{year}{2013}\natexlab{}.
\newblock \showarticletitle{{INSPIRE}: The insieme parallel intermediate
  representation}. In \bibinfo{booktitle}{\emph{Proceedings of the 22nd
  International Conference on Parallel Architectures and Compilation
  Techniques}}. \bibinfo{pages}{7--17}.
\newblock
\urldef\tempurl%
\url{https://doi.org/10.1109/PACT.2013.6618799}
\showDOI{\tempurl}


\bibitem[Klusik et~al\mbox{.}(2001)]%
        {Klusik_Loogen_Priebe_Rubio_2001}
\bibfield{author}{\bibinfo{person}{Ulrike Klusik}, \bibinfo{person}{Rita
  Loogen}, \bibinfo{person}{Steffen Priebe}, {and} \bibinfo{person}{Fernando
  Rubio}.} \bibinfo{year}{2001}\natexlab{}.
\newblock \showarticletitle{Implementation Skeletons in Eden: Low-Effort
  Parallel Programming} \emph{(\bibinfo{series}{Lecture Notes in Computer
  Science})}, \bibfield{editor}{\bibinfo{person}{Markus Mohnen} {and}
  \bibinfo{person}{Pieter Koopman}} (Eds.). \bibinfo{publisher}{Springer},
  \bibinfo{address}{Berlin, Heidelberg}, \bibinfo{pages}{71–88}.
\newblock
\showISBNx{978-3-540-45361-1}
\urldef\tempurl%
\url{https://doi.org/10.1007/3-540-45361-X_5}
\showDOI{\tempurl}


\bibitem[Kotsifakou et~al\mbox{.}(2018)]%
        {hpvm}
\bibfield{author}{\bibinfo{person}{Maria Kotsifakou}, \bibinfo{person}{Prakalp
  Srivastava}, \bibinfo{person}{Matthew~D. Sinclair}, \bibinfo{person}{Rakesh
  Komuravelli}, \bibinfo{person}{Vikram Adve}, {and} \bibinfo{person}{Sarita
  Adve}.} \bibinfo{year}{2018}\natexlab{}.
\newblock \showarticletitle{{HPVM:} Heterogeneous Parallel Virtual Machine}. In
  \bibinfo{booktitle}{\emph{Proceedings of the 23rd ACM SIGPLAN Symposium on
  Principles and Practice of Parallel Programming}} (Vienna, Austria)
  \emph{(\bibinfo{series}{PPoPP '18})}. \bibinfo{publisher}{Association for
  Computing Machinery}, \bibinfo{address}{New York, NY, USA},
  \bibinfo{pages}{68–80}.
\newblock
\showISBNx{9781450349826}
\urldef\tempurl%
\url{https://doi.org/10.1145/3178487.3178493}
\showDOI{\tempurl}


\bibitem[Kulkarni et~al\mbox{.}(2007)]%
        {galois}
\bibfield{author}{\bibinfo{person}{Milind Kulkarni}, \bibinfo{person}{Keshav
  Pingali}, \bibinfo{person}{Bruce Walter}, \bibinfo{person}{Ganesh
  Ramanarayanan}, \bibinfo{person}{Kavita Bala}, {and} \bibinfo{person}{L.~Paul
  Chew}.} \bibinfo{year}{2007}\natexlab{}.
\newblock \showarticletitle{Optimistic Parallelism Requires Abstractions}. In
  \bibinfo{booktitle}{\emph{Proceedings of the 28th ACM SIGPLAN Conference on
  Programming Language Design and Implementation}} (San Diego, California, USA)
  \emph{(\bibinfo{series}{PLDI '07})}. \bibinfo{publisher}{Association for
  Computing Machinery}, \bibinfo{address}{New York, NY, USA},
  \bibinfo{pages}{211–222}.
\newblock
\showISBNx{9781595936332}
\urldef\tempurl%
\url{https://doi.org/10.1145/1250734.1250759}
\showDOI{\tempurl}


\bibitem[Lam(1988)]%
        {lam1988software}
\bibfield{author}{\bibinfo{person}{Monica Lam}.}
  \bibinfo{year}{1988}\natexlab{}.
\newblock \showarticletitle{Software pipelining: An effective scheduling
  technique for VLIW machines}. In \bibinfo{booktitle}{\emph{Proceedings of the
  ACM SIGPLAN 1988 conference on Programming Language design and
  Implementation}}. \bibinfo{pages}{318--328}.
\newblock


\bibitem[Lattner and Adve(2004)]%
        {llvm}
\bibfield{author}{\bibinfo{person}{Chris Lattner} {and} \bibinfo{person}{Vikram
  Adve}.} \bibinfo{year}{2004}\natexlab{}.
\newblock \showarticletitle{{LLVM:} A compilation framework for lifelong
  program analysis \& transformation}. In \bibinfo{booktitle}{\emph{Proceedings
  of the international symposium on Code generation and optimization:
  feedback-directed and runtime optimization}}. IEEE Computer Society,
  \bibinfo{pages}{75}.
\newblock


\bibitem[Lee et~al\mbox{.}(2000)]%
        {lee2000compiler}
\bibfield{author}{\bibinfo{person}{Chingren Lee}, \bibinfo{person}{Jenq~Kuen
  Lee}, {and} \bibinfo{person}{TingTing Hwang}.}
  \bibinfo{year}{2000}\natexlab{}.
\newblock \showarticletitle{Compiler optimization on instruction scheduling for
  low power}. In \bibinfo{booktitle}{\emph{Proceedings 13th International
  Symposium on System Synthesis}}. IEEE, \bibinfo{pages}{55--60}.
\newblock


\bibitem[Li et~al\mbox{.}(2007)]%
        {li_lightweight_2007}
\bibfield{author}{\bibinfo{person}{Peng Li}, \bibinfo{person}{Simon Marlow},
  \bibinfo{person}{Simon Peyton~Jones}, {and} \bibinfo{person}{Andrew
  Tolmach}.} \bibinfo{year}{2007}\natexlab{}.
\newblock \showarticletitle{Lightweight concurrency primitives for {GHC}}. In
  \bibinfo{booktitle}{\emph{Proceedings of the {ACM} {SIGPLAN} workshop on
  {Haskell} workshop}} \emph{(\bibinfo{series}{Haskell '07})}.
  \bibinfo{publisher}{Association for Computing Machinery},
  \bibinfo{address}{New York, NY, USA}, \bibinfo{pages}{107--118}.
\newblock
\showISBNx{978-1-59593-674-5}
\urldef\tempurl%
\url{https://doi.org/10.1145/1291201.1291217}
\showDOI{\tempurl}


\bibitem[Ma et~al\mbox{.}(2021)]%
        {DBLP:conf/sc/MaWNCHLHCHD21}
\bibfield{author}{\bibinfo{person}{Jiacheng Ma}, \bibinfo{person}{Wenyi Wang},
  \bibinfo{person}{Aaron Nelson}, \bibinfo{person}{Michael Cuevas},
  \bibinfo{person}{Brian Homerding}, \bibinfo{person}{Conghao Liu},
  \bibinfo{person}{Zhen Huang}, \bibinfo{person}{Simone Campanoni},
  \bibinfo{person}{Kyle~C. Hale}, {and} \bibinfo{person}{Peter~A. Dinda}.}
  \bibinfo{year}{2021}\natexlab{}.
\newblock \showarticletitle{Paths to OpenMP in the kernel}. In
  \bibinfo{booktitle}{\emph{{SC} '21: The International Conference for High
  Performance Computing, Networking, Storage and Analysis, St. Louis, Missouri,
  USA, November 14 - 19, 2021}}, \bibfield{editor}{\bibinfo{person}{Bronis~R.
  de~Supinski}, \bibinfo{person}{Mary~W. Hall}, {and} \bibinfo{person}{Todd
  Gamblin}} (Eds.). \bibinfo{publisher}{{ACM}}, \bibinfo{pages}{65:1--65:17}.
\newblock
\urldef\tempurl%
\url{https://doi.org/10.1145/3458817.3476183}
\showDOI{\tempurl}


\bibitem[Marlow(2012)]%
        {marlow_parallel_2012}
\bibfield{author}{\bibinfo{person}{Simon Marlow}.}
  \bibinfo{year}{2012}\natexlab{}.
\newblock \showarticletitle{Parallel and Concurrent Programming in Haskell}.
\newblock In \bibinfo{booktitle}{\emph{Central {European} {Functional}
  {Programming} {School}: 4th {Summer} {School}, {CEFP} 2011, {Budapest},
  {Hungary}, {June} 14-24, 2011, {Revised} {Selected} {Papers}}},
  \bibfield{editor}{\bibinfo{person}{Viktória Zsók}, \bibinfo{person}{Zoltán
  Horváth}, {and} \bibinfo{person}{Rinus Plasmeijer}} (Eds.).
  \bibinfo{publisher}{Springer}, \bibinfo{address}{Berlin, Heidelberg},
  \bibinfo{pages}{339--401}.
\newblock
\showISBNx{978-3-642-32096-5}
\urldef\tempurl%
\url{https://doi.org/10.1007/978-3-642-32096-5\_7}
\showDOI{\tempurl}


\bibitem[Matni et~al\mbox{.}(2022)]%
        {matni2022noelle}
\bibfield{author}{\bibinfo{person}{Angelo Matni},
  \bibinfo{person}{Enrico~Armenio Deiana}, \bibinfo{person}{Yian Su},
  \bibinfo{person}{Lukas Gross}, \bibinfo{person}{Souradip Ghosh},
  \bibinfo{person}{Sotiris Apostolakis}, \bibinfo{person}{Ziyang Xu},
  \bibinfo{person}{Zujun Tan}, \bibinfo{person}{Ishita Chaturvedi},
  \bibinfo{person}{Brian Homerding}, \bibinfo{person}{Tommy McMichen},
  \bibinfo{person}{David~I. August}, {and} \bibinfo{person}{Simone Campanoni}.}
  \bibinfo{year}{2022}\natexlab{}.
\newblock \showarticletitle{{NOELLE} Offers Empowering {LLVM} Extensions}. In
  \bibinfo{booktitle}{\emph{Proceedings of the 20th IEEE/ACM International
  Symposium on Code Generation and Optimization}} (Virtual Event, Republic of
  Korea) \emph{(\bibinfo{series}{CGO '22})}. \bibinfo{publisher}{IEEE Press},
  \bibinfo{pages}{179–192}.
\newblock
\showISBNx{9781665405843}
\urldef\tempurl%
\url{https://doi.org/10.1109/CGO53902.2022.9741276}
\showDOI{\tempurl}


\bibitem[Nandivada et~al\mbox{.}(2013)]%
        {psg}
\bibfield{author}{\bibinfo{person}{V.~Krishna Nandivada}, \bibinfo{person}{Jun
  Shirako}, \bibinfo{person}{Jisheng Zhao}, {and} \bibinfo{person}{Vivek
  Sarkar}.} \bibinfo{year}{2013}\natexlab{}.
\newblock \showarticletitle{A Transformation Framework for Optimizing
  Task-Parallel Programs}.
\newblock \bibinfo{journal}{\emph{ACM Trans. Program. Lang. Syst.}}
  \bibinfo{volume}{35}, \bibinfo{number}{1}, Article \bibinfo{articleno}{3}
  (\bibinfo{date}{apr} \bibinfo{year}{2013}), \bibinfo{numpages}{48}~pages.
\newblock
\showISSN{0164-0925}
\urldef\tempurl%
\url{https://doi.org/10.1145/2450136.2450138}
\showDOI{\tempurl}


\bibitem[Ngo et~al\mbox{.}(1997)]%
        {pp_data_par_lang}
\bibfield{author}{\bibinfo{person}{Ton Ngo}, \bibinfo{person}{Lawrence Snyder},
  {and} \bibinfo{person}{Bradford Chamberlain}.}
  \bibinfo{year}{1997}\natexlab{}.
\newblock \showarticletitle{Portable Performance of Data Parallel Languages}.
  In \bibinfo{booktitle}{\emph{Proceedings of the 1997 ACM/IEEE Conference on
  Supercomputing}} (San Jose, CA) \emph{(\bibinfo{series}{SC '97})}.
  \bibinfo{publisher}{Association for Computing Machinery},
  \bibinfo{address}{New York, NY, USA}, \bibinfo{pages}{1–20}.
\newblock
\showISBNx{0897919858}
\urldef\tempurl%
\url{https://doi.org/10.1145/509593.509611}
\showDOI{\tempurl}


\bibitem[{OpenMP Architecture Review Board}(2018)]%
        {openmp50}
\bibfield{author}{\bibinfo{person}{{OpenMP Architecture Review Board}}.}
  \bibinfo{year}{2018}\natexlab{}.
\newblock \bibinfo{title}{{OpenMP} Application Program Interface Version 5.0}.
\newblock
\newblock
\urldef\tempurl%
\url{https://www.openmp.org/wp-content/uploads/OpenMP-API-Specification-5.0.pdf}
\showURL{%
\tempurl}


\bibitem[Peyton~Jones et~al\mbox{.}(2008)]%
        {peyton_jones_harnessing_2008}
\bibfield{author}{\bibinfo{person}{Simon Peyton~Jones}, \bibinfo{person}{Roman
  Leshchinskiy}, \bibinfo{person}{Gabriele Keller}, {and}
  \bibinfo{person}{Manuel Chakravarty}.} \bibinfo{year}{2008}\natexlab{}.
\newblock \showarticletitle{Harnessing the Multicores: Nested Data Parallelism
  in {Haskell}}.
\newblock \bibinfo{journal}{\emph{Leibniz International Proceedings in
  Informatics, LIPIcs}}  \bibinfo{volume}{2} (\bibinfo{date}{Dec.}
  \bibinfo{year}{2008}).
\newblock
\showISSN{978-3-540-89329-5}
\urldef\tempurl%
\url{https://doi.org/10.4230/LIPIcs.FSTTCS.2008.1769}
\showDOI{\tempurl}


\bibitem[Phothilimthana et~al\mbox{.}(2013)]%
        {pp_herterogeneous_arch}
\bibfield{author}{\bibinfo{person}{Phitchaya~Mangpo Phothilimthana},
  \bibinfo{person}{Jason Ansel}, \bibinfo{person}{Jonathan Ragan-Kelley}, {and}
  \bibinfo{person}{Saman Amarasinghe}.} \bibinfo{year}{2013}\natexlab{}.
\newblock \showarticletitle{Portable Performance on Heterogeneous
  Architectures}. In \bibinfo{booktitle}{\emph{Proceedings of the Eighteenth
  International Conference on Architectural Support for Programming Languages
  and Operating Systems}} (Houston, Texas, USA) \emph{(\bibinfo{series}{ASPLOS
  '13})}. \bibinfo{publisher}{Association for Computing Machinery},
  \bibinfo{address}{New York, NY, USA}, \bibinfo{pages}{431–444}.
\newblock
\showISBNx{9781450318709}
\urldef\tempurl%
\url{https://doi.org/10.1145/2451116.2451162}
\showDOI{\tempurl}


\bibitem[Raghunathan et~al\mbox{.}(2016)]%
        {raghunathan_hierarchical_2016}
\bibfield{author}{\bibinfo{person}{Ram Raghunathan}, \bibinfo{person}{Stefan~K.
  Muller}, \bibinfo{person}{Umut~A. Acar}, {and} \bibinfo{person}{Guy
  Blelloch}.} \bibinfo{year}{2016}\natexlab{}.
\newblock \showarticletitle{Hierarchical memory management for parallel
  programs}. In \bibinfo{booktitle}{\emph{Proceedings of the 21st {ACM}
  {SIGPLAN} {International} {Conference} on {Functional} {Programming}}}
  \emph{(\bibinfo{series}{{ICFP} 2016})}. \bibinfo{publisher}{Association for
  Computing Machinery}, \bibinfo{address}{New York, NY, USA},
  \bibinfo{pages}{392--406}.
\newblock
\showISBNx{978-1-4503-4219-3}
\urldef\tempurl%
\url{https://doi.org/10.1145/2951913.2951935}
\showDOI{\tempurl}


\bibitem[Reddi et~al\mbox{.}(2010)]%
        {Reddi:2010:EVE:1839667.1839674}
\bibfield{author}{\bibinfo{person}{Vijay~Janapa Reddi}, \bibinfo{person}{Simone
  Campanoni}, \bibinfo{person}{Meeta~S. Gupta}, \bibinfo{person}{Michael~D.
  Smith}, \bibinfo{person}{Gu-Yeon Wei}, \bibinfo{person}{David Brooks}, {and}
  \bibinfo{person}{Kim Hazelwood}.} \bibinfo{year}{2010}\natexlab{}.
\newblock \showarticletitle{Eliminating Voltage Emergencies via Software-guided
  Code Transformations}.
\newblock \bibinfo{journal}{\emph{ACM Trans. Archit. Code Optim.}}
  \bibinfo{volume}{7}, \bibinfo{number}{2}, Article \bibinfo{articleno}{12}
  (\bibinfo{date}{Oct.} \bibinfo{year}{2010}), \bibinfo{numpages}{28}~pages.
\newblock
\showISSN{1544-3566}
\urldef\tempurl%
\url{https://doi.org/10.1145/1839667.1839674}
\showDOI{\tempurl}


\bibitem[Sarkar(1989)]%
        {sarkar_thesis}
\bibfield{author}{\bibinfo{person}{Vivek Sarkar}.}
  \bibinfo{year}{1989}\natexlab{}.
\newblock \bibinfo{booktitle}{\emph{Partitioning and Scheduling Parallel
  Programs for Multiprocessors}}.
\newblock \bibinfo{publisher}{MIT Press}, \bibinfo{address}{Cambridge, MA,
  USA}.
\newblock
\showISBNx{0262691302}


\bibitem[Sarkar(1997)]%
        {sarkar_opt_ppg}
\bibfield{author}{\bibinfo{person}{Vivek Sarkar}.}
  \bibinfo{year}{1997}\natexlab{}.
\newblock \showarticletitle{Analysis and Optimization of Explicitly Parallel
  Programs Using the Parallel Program Graph Representation}. In
  \bibinfo{booktitle}{\emph{LCPC}}.
\newblock


\bibitem[Sarkar and Simons(1993)]%
        {ppg}
\bibfield{author}{\bibinfo{person}{Vivek Sarkar} {and} \bibinfo{person}{Barbara
  Simons}.} \bibinfo{year}{1993}\natexlab{}.
\newblock \showarticletitle{Parallel Program Graphs and Their Classification}.
  In \bibinfo{booktitle}{\emph{Proceedings of the 6th International Workshop on
  Languages and Compilers for Parallel Computing}}.
  \bibinfo{publisher}{Springer-Verlag}, \bibinfo{address}{Berlin, Heidelberg},
  \bibinfo{pages}{633–655}.
\newblock
\showISBNx{3540576592}


\bibitem[Schardl et~al\mbox{.}(2017)]%
        {tapir}
\bibfield{author}{\bibinfo{person}{Tao~B. Schardl}, \bibinfo{person}{William~S.
  Moses}, {and} \bibinfo{person}{Charles~E. Leiserson}.}
  \bibinfo{year}{2017}\natexlab{}.
\newblock \showarticletitle{Tapir: Embedding Fork-Join Parallelism into
  {LLVM}'s Intermediate Representation}. In
  \bibinfo{booktitle}{\emph{Proceedings of the 22nd ACM SIGPLAN Symposium on
  Principles and Practice of Parallel Programming}} (Austin, Texas, USA)
  \emph{(\bibinfo{series}{PPoPP '17})}. \bibinfo{publisher}{Association for
  Computing Machinery}, \bibinfo{address}{New York, NY, USA},
  \bibinfo{pages}{249–265}.
\newblock
\showISBNx{9781450344937}
\urldef\tempurl%
\url{https://doi.org/10.1145/3018743.3018758}
\showDOI{\tempurl}


\bibitem[Sivaramakrishnan et~al\mbox{.}(2014)]%
        {sivaramakrishnan_multimlton_2014}
\bibfield{author}{\bibinfo{person}{K.~C. Sivaramakrishnan},
  \bibinfo{person}{Lukasz Ziarek}, {and} \bibinfo{person}{Suresh Jagannathan}.}
  \bibinfo{year}{2014}\natexlab{}.
\newblock \showarticletitle{{MultiMLton}: {A} multicore-aware runtime for
  standard {ML}}.
\newblock \bibinfo{journal}{\emph{Journal of Functional Programming}}
  \bibinfo{volume}{24}, \bibinfo{number}{6} (\bibinfo{date}{Nov.}
  \bibinfo{year}{2014}), \bibinfo{pages}{613--674}.
\newblock
\showISSN{0956-7968, 1469-7653}
\urldef\tempurl%
\url{https://doi.org/10.1017/S0956796814000161}
\showDOI{\tempurl}
\newblock
\shownote{Publisher: Cambridge University Press}.


\bibitem[Srinivasan and Wolfe(1991)]%
        {wolfe_pcfg}
\bibfield{author}{\bibinfo{person}{Harini Srinivasan} {and}
  \bibinfo{person}{Michael Wolfe}.} \bibinfo{year}{1991}\natexlab{}.
\newblock \showarticletitle{Analyzing Programs with Explicit Parallelism}. In
  \bibinfo{booktitle}{\emph{Proceedings of the Fourth International Workshop on
  Languages and Compilers for Parallel Computing}}.
  \bibinfo{publisher}{Springer-Verlag}, \bibinfo{address}{Berlin, Heidelberg},
  \bibinfo{pages}{405–419}.
\newblock
\showISBNx{354055422X}


\bibitem[Tran et~al\mbox{.}(2017)]%
        {tran2017clairvoyance}
\bibfield{author}{\bibinfo{person}{Kim-Anh Tran}, \bibinfo{person}{Trevor~E
  Carlson}, \bibinfo{person}{Konstantinos Koukos}, \bibinfo{person}{Magnus
  Sj{\"a}lander}, \bibinfo{person}{Vasileios Spiliopoulos},
  \bibinfo{person}{Stefanos Kaxiras}, {and} \bibinfo{person}{Alexandra
  Jimborean}.} \bibinfo{year}{2017}\natexlab{}.
\newblock \showarticletitle{Clairvoyance: Look-ahead compile-time scheduling}.
  In \bibinfo{booktitle}{\emph{2017 IEEE/ACM International Symposium on Code
  Generation and Optimization (CGO)}}. IEEE, \bibinfo{pages}{171--184}.
\newblock


\bibitem[Vachharajani et~al\mbox{.}(2007)]%
        {dswp}
\bibfield{author}{\bibinfo{person}{Neil Vachharajani}, \bibinfo{person}{Ram
  Rangan}, \bibinfo{person}{Easwaran Raman}, \bibinfo{person}{Matthew~J
  Bridges}, \bibinfo{person}{Guilherme Ottoni}, {and} \bibinfo{person}{David~I
  August}.} \bibinfo{year}{2007}\natexlab{}.
\newblock \showarticletitle{Speculative decoupled software pipelining}. In
  \bibinfo{booktitle}{\emph{16th International Conference on Parallel
  Architecture and Compilation Techniques (PACT 2007)}}. IEEE,
  \bibinfo{pages}{49--59}.
\newblock


\bibitem[Westrick et~al\mbox{.}(2020)]%
        {westrick_disentanglement_2020}
\bibfield{author}{\bibinfo{person}{Sam Westrick}, \bibinfo{person}{Rohan
  Yadav}, \bibinfo{person}{Matthew Fluet}, {and} \bibinfo{person}{Umut~A.
  Acar}.} \bibinfo{year}{2020}\natexlab{}.
\newblock \showarticletitle{Disentanglement in nested-parallel programs}.
\newblock \bibinfo{journal}{\emph{Proceedings of the ACM on Programming
  Languages}} \bibinfo{volume}{4}, \bibinfo{number}{POPL} (\bibinfo{date}{Jan.}
  \bibinfo{year}{2020}), \bibinfo{pages}{1--32}.
\newblock
\showISSN{2475-1421}
\urldef\tempurl%
\url{https://doi.org/10.1145/3371115}
\showDOI{\tempurl}


\bibitem[Zhang et~al\mbox{.}(2021)]%
        {iiswc}
\bibfield{author}{\bibinfo{person}{Xiaochun Zhang}, \bibinfo{person}{Timothy~M.
  Jones}, {and} \bibinfo{person}{Simone Campanoni}.}
  \bibinfo{year}{2021}\natexlab{}.
\newblock \showarticletitle{Quantifying the Semantic Gap Between Serial and
  Parallel Programming}. In \bibinfo{booktitle}{\emph{2021 IEEE International
  Symposium on Workload Characterization (IISWC)}}. \bibinfo{pages}{151--162}.
\newblock
\urldef\tempurl%
\url{https://doi.org/10.1109/IISWC53511.2021.00024}
\showDOI{\tempurl}


\end{thebibliography}
\newpage
\pagebreak
\clearpage
\appendix
\section{Sufficiency of PS-PDG for Cilk} \label{sec:appendix}
This Appendix describes how to map the Cilk parallel programming model to PS-PDG.
In particular, we refer to OpenCilk 2.0 \cite{opencilk} as Cilk as it is the only actively maintained implementation as of this writing.
Following OpenCilk 2.0, we exclude features such as inlets, array operations, elemental functions, and the \code{simd} pragma.
Note that inlets were removed as early as Intel Cilk Plus \cite{cilkplus_spec}.
Additionally, since array operations and elemental functions are language constructs to exploit data parallelism, their semantics can be equivalently expressed as a \code{cilk\_for} if desired.
Similar to OpenMP, we do not consider Cilk clauses that control the amount of parallelism to generate such as \code{grainsize}.
Note however that the Cilk \code{simd} is semantically identical to the OpenMP \code{simd}.

The execution model of Cilk is expressed in the PS-PDG as follows.
The \code{cilk\_spawn} construct is represented as a hierarchical single-entry single-exit (SESE) node that contains two nodes: an inner entry node (referred to as a knot in the Cilk community) with two directed outgoing edges: one to a node within the hierarchical node representing the function call within the spawn construct, and the other exiting the hierarchical node.
Each directed outgoing edge from the knot represents an outgoing strand of execution (thread) from the spawn knot.
That is, the node representing the spawned function call represents forking a thread to call that function to be joined at the next synchronization point.
Similar to \code{omp barrier}, \code{cilk\_sync} is represented by a node with incoming edges from all nodes that contain spawned function calls in the smallest hierarchical node that contains it.
The construct \code{cilk\_scope} is represented by a SESE hierarchical node that contains a \code{cilk\_sync} as the exit node and contains the nodes needed to represent the contents of the Cilk scope.
The \code{cilk\_sync} node is necessary as there is an implicit synchronization at the end of every Cilk scope.
Finally, \code{cilk\_for} is represented identically to \code{omp parallel for}.

Cilk hyperobjects are represented as reducible parallel semantic variables.
Recall that Cilk hyperobjects are copied from the parent function into the child strand, and after the spawned child is joined back into the parent, the parent's view of the object is reduced with the child's view using the given reducer operation by the programmer.
At this point, the child's view of the object is destroyed along with the spawned strand.
By design, this semantic also provides support for other hyperobjects such as holders, which are a special case of reducers.
For these classes of hyperobjects, parallel semantic variables and their properties provide the relevant semantics for Cilk.
Therefore, we conclude that the PS-PDG abstraction possesses all the features needed to capture the Cilk programming model, as well as OpenMP.

\end{document}